\documentclass[twocolumn]{aastex701}
\usepackage[utf8]{inputenc}
\usepackage{caption}
\usepackage{subcaption}
\usepackage{colortbl}
\usepackage{verbatim}
\usepackage{amsmath}
\usepackage{nicefrac}

\shorttitle{Post-Algols}
\shortauthors{Frank et al.}

\newcommand{\cs}{$\chi^2$\,}
\newcommand{\ks}{\,km\,s$^{-1}$\,}
\newcommand{\Lsun}{L$_\odot$}
\newcommand{\Rsun}{R$_\odot$}
\newcommand{\Msun}{M$_\odot$}
\newcommand{\vsini}{$v$\,sin\,$i$ }

\begin{document}

\title{Detached Post-Algol Eclipsing Binaries Caught Between Case A and Case AB Mass Transfer}

\author[0009-0002-2866-9788]{Megan G. Frank}
\affil{Department of Physics and Astronomy, University of Wyoming, 1000~E.~University~Ave., Dept.~3905, Laramie, WY 82071, USA}
\email{mfrank13@uwyo.edu}

\author[0000-0002-0870-6388]{Maxwell Moe}
\affil{Department of Physics and Astronomy, University of Wyoming, 1000~E.~University~Ave., Dept.~3905, Laramie, WY 82071, USA}
\email{mmoe2@uwyo.edu}

\author[0000-0001-5510-2424]{Nathan Smith}
\affil{Steward Observatory, University of Arizona, 933~N.~Cherry~Ave.,~Tucson,~AZ 85721,~USA}
\email{nathans@as.arizona.edu}

\keywords{binary stars: close, eclipsing; stars: evolved, evolution, early-type}

\begin{abstract}
For sixty years, stellar evolutionary models have predicted that intermediate-mass stars slightly contract on the terminal-age main-sequence (TAMS) as they exhaust hydrogen in their convective cores, producing the main-sequence (MS) hook on the Hertzsprung-Russell diagram. Contraction along the TAMS has not previously been observationally verified, but an evolved eclipsing binary (EB) with a component on the TAMS can test this prediction. In a very close binary with an orbital period of less than a week, the primary star initially fills its Roche lobe on the MS (Case A mass transfer), and the binary can invert mass ratios, producing a classical Algol. The subgiant donor then contracts on the TAMS and detaches slightly from its Roche lobe. The subgiant subsequently re-expands and refills its Roche lobe as it evolves toward the Hertzsprung Gap (Case AB mass transfer). We report the discovery of detached post-Algol EB candidates in the LMC caught between Case~A and Case~AB mass transfer. Their OGLE light curves feature strong reflection effects as the hot primary (former mass gainer) irradiates the cool subgiant secondary. We analyze multi-epoch echelle spectra of four post-Algol candidates taken with the MIKE spectrograph at the 6.5m Magellan-Clay telescope. The primaries have mid-B MS atmospheres ($M_1$ = 6\,-\,8\,\Msun). We measure dynamical masses of the subgiant secondaries to be $M_2$ = 0.9\,-\,1.2\,\Msun. Detailed fitting of the OGLE light curves with PHOEBE reveals that the subgiants have Roche lobe fill factors of $RLFF_2$ = 73\%\,-\,89\%, consistent with binary evolution models. Our discovery of detached post-Algol candidates provides the first empirical evidence that intermediate-mass stars contract along the TAMS. 
\end{abstract}

\section{Introduction}\label{sec:Intro}

Most stars are born in binary or multiple systems (see \citet{Offner2023} for a recent review). Large-scale photometric time-domain surveys such as the Optical Gravitational Lensing Experiment (OGLE) have discovered hundreds of thousands of eclipsing binaries (EBs) in our Milky Way \citep{Soszynski2016} and tens of thousands in the nearby Magellanic Clouds \citep{Pawlak2016,Glowacki2024}. 
The third data release of {\it Gaia} contains more than 2 million EBs \citep{Mowlavi2023}, and the Rubin Observatory's survey is expected to identify 24 million EBs \citep{Prsa2011}. 

Such large samples contain interesting EBs in short-lived phases of evolution that exhibit unique characteristics in their light curves. Some EBs feature sinusoidal reflection effects between eclipses due to a hot star irradiating the near side of a cool star. For example, HW~Vir is the prototype of reflecting post-common-envelope EBs composed of a hot B~subdwarf that irradiates a cool M~dwarf in a tight orbit of 2.8 hours \citep{BrownSevilla2021}. The heated day side of the M dwarf is hotter and more luminous than its night side, yielding a sinusoidal variation in the light curve as the heated side comes in and out of view during its orbit. To produce reflection effect amplitudes greater than a few percent, the hot component must be similar in size to the cooler component but at least three times the temperature \citep{Moe2015a}. Hence an EB with reflection effects strong enough to be detectable with ground-based photometry cannot consist of two garden-variety main-sequence (MS) stars. 

By searching the OGLE-III database of EBs in the LMC \citep{Gracyk2011}, \citet{Moe2015a} identified 18 detached EBs with B-type primaries, reflection effect amplitudes $\Delta I_{\rm Refl}$~=~3\%\,-\,12\%, and orbital periods $P$ = 3\,-\,8 days, the majority of which reside in bright H\,II regions. They concluded that these reflecting systems are nascent EBs whereby a young B-type MS star irradiates a cool, large, solar-type pre-MS companion that is still contracting toward the zero-age MS. Such nascent reflecting EBs with extreme mass ratios have since been discovered in star-forming regions within our own Milky Way \citep{Jerzykiewicz2021,Pigulski2024, Naze2025}.  Assuming that the two components are coeval and evolving along their respective single-star tracks, \citet{Moe2015a} fit models to the OGLE-III EB light curves in order to measure their physical properties. EBs with larger pre-MS companions are younger, exhibit larger reflection effect amplitudes, and typically reside in brighter and more compact H\,II regions.

The notable exception is OGLE-LMC-ECL-05898, hereafter EB-5898. EB-5898 exhibits a large reflection effect amplitude of $\Delta I_{\rm Refl}$~=~10\% but does not reside in a bright or compact H\,II region \citep{Moe2015a}. The assumption that the EB is composed of a B-type MS primary and pre-MS secondary resulted in a rather poor light curve fit with \cs = 1.27 and probability to exceed of $p$~$<$~0.001. The light curve fit to EB-5898 is considerably the worst of the sample (see Table 2 in \citealt{Moe2015a}). Moreover, the inferred radius and luminosity of the supposed pre-MS secondary in EB-5898 are substantially greater than the other systems, placing it in an extremely short-lived and therefore highly unlikely isolated location on the Hertzsprung-Russell diagram (HRD; see Fig.~6 in \citealt{Moe2015a}). Finally, the measured dust extinction toward EB-5898 based on the joint fits to the OGLE-III I-band and V-band light curves is highly inconsistent with dust extinction values measured via more robust techniques (see Fig.~7 in \citealt{Moe2015a}). Clearly, EB-5898 does not belong to the class of nascent EBs with pre-MS secondaries.

In the present study, we instead conclude that EB-5898 is a post-Algol containing a B-type MS primary that was the former accretor and a cool subgiant secondary that was previously the donor but has recently detached from its Roche lobe. In Section~\ref{sec:post-Algol}, we discuss Algol evolution and the formation of detached post-Algols like EB-5898. In Section~\ref{sec:Obj}, we search the OGLE-III database of LMC EBs for additional reflecting EBs that do not reside in H\,II regions. We obtain high-resolution, multi-epoch spectra of four post-Algol candidates and discuss the data reduction in Section~\ref{sec:Obs}. We fit atmospheric models to the spectra, measure radial velocity curves, and fit physical models to the light curves in Sections~\ref{sec:PrimParams}, \ref{sec:RVcurves}, and \ref{sec:LCfitting}, respectively. We present and discuss our final combined results in Section~\ref{sec:Results}. We summarize our main conclusions in Section~\ref{sec:summary}.

\section{Post-Algols}
\label{sec:post-Algol}

\begin{figure}[t!]
    \centering
    \includegraphics[width=\linewidth]{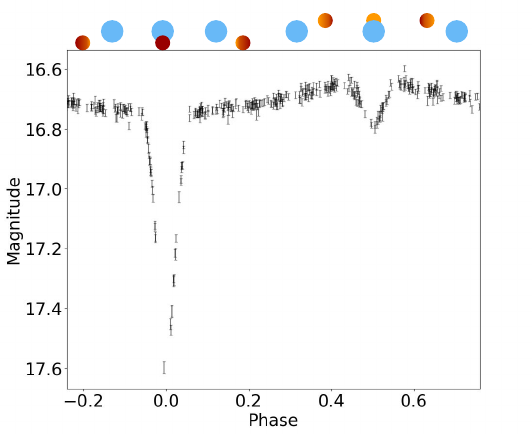}
    \caption{OGLE-III I-band light curve of our prototype post-Algol EB-5898. A schematic diagram of the positions of the B-type MS primary (blue) and irradiated subgiant (red) are displayed at the top as a function of orbital phase. }
    \label{fig:lcwithPhases}
\end{figure}

We show the OGLE-III I-band light curve of EB-5898 in Fig.~\ref{fig:lcwithPhases}. This EB features a deep primary eclipse of $\Delta I_1$~=~0.8~mag and a prominent reflection effect with amplitude $\Delta I_{\rm Refl}$~=~0.1~mag. The eclipses are relatively narrow and there is no ellipsoidal variability between eclipses, both indicating that the EB is detached (see also Section~\ref{sec:Obj}).

Intermediate-mass stars near the terminal-age MS (TAMS) shrink by 10\%\,-\,20\% as the convective core contracts and becomes semi-degenerate \citep{Iben1967a,Iben1967b}. According to MIST evolutionary tracks, a 5\,\Msun\ star attains a maximum radius of 6.2\,\Rsun\ on the MS, increases in effective temperature and luminosity while contracting to 5.4\,\Rsun\ during the TAMS, and subsequently expands and cools again along the Hertzsprung Gap \citep{Choi2016}. The TAMS hook is a common feature of evolutionary tracks of intermediate-mass stars on the HRD \citep{Dotter2008,Bressan2012,Choi2016}. Despite the ubiquity of the TAMS hook in theoretical evolutionary tracks, there has hitherto been little observational evidence that intermediate-mass stars contract on the TAMS \citep{deBurgos2025}.

\begin{figure}[t!]
    \centering
    \includegraphics[width=\linewidth]{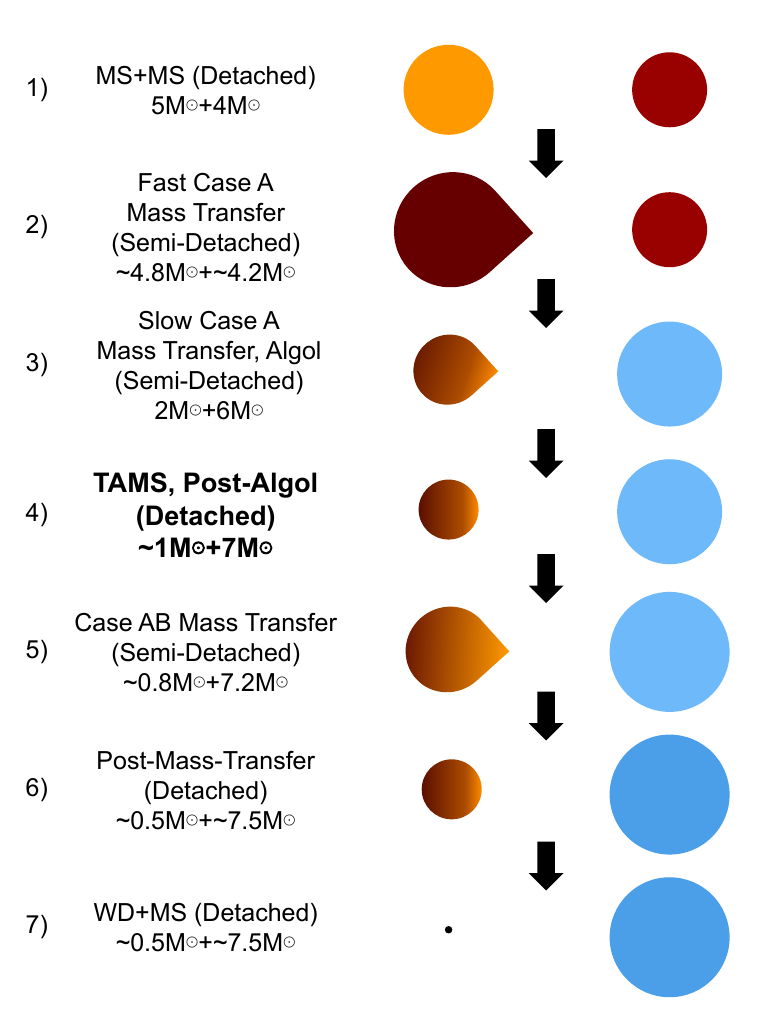}
    \caption{Evolutionary sequence of intermediate-mass binaries, highlighting the post-Algol phase between Case A and AB mass transfer. }
    \label{fig:evTrack}
\end{figure}

For close binary stars that first interact on the MS, there can be two phases of mass transfer, Case A and AB, that are separated when the donor star shrinks on the TAMS. We display a cartoon illustration of close binary evolution that features this intermediate stage of evolution in Fig.~\ref{fig:evTrack}. Detailed physical models of Case A and AB mass transfer are presented in Fig.~3 of \citet{Pols1994}, Fig.~2 of \citet{Wellstein2001}, Fig.~4 of \citet{Siess2018}, and Fig.~2 of \citet{Sen2022}, albeit at higher masses than our systems considered here.  Case A and AB mass transfer are further discussed in other binary population synthesis studies \citep{Petrovic2005,Schurmann2024}. Given two very close MS stars with an orbital period of less than a week (phase 1 in Fig.~\ref{fig:evTrack}), the primary fills its Roche lobe while still on the MS. Mass transfer initially ensues on a rapid thermal timescale, i.e., fast Case A mass transfer (phase 2), and the mass ratio reverses as the system stabilizes. Mass transfer then proceeds on a longer nuclear timescale, i.e., slow Case A mass transfer (phase 3), where classical Algols are observed. During this phase, the accretor is a B-type MS star while the less massive donor is a cool, Roche-lobe filling subgiant. As the subgiant donor evolves toward the TAMS, it shrinks and detaches from its Roche lobe (phase 4), a stage that we designate post-Algol. Just as single intermediate-mass stars contract on the TAMS, the donor star exhausts hydrogen in its core on the TAMS, the core contracts, and thus the entire donor star also contracts \citep{Pols1994,Schurmann2024}. According to the binary evolution models of \citet{Sen2022}, the subgiant shrinks to a Roche lobe fill factor of RLFF = 75\% during this post-Algol phase. At the end of the TAMS, if there is sufficient envelope mass and the orbit has not substantially widened, the subgiant re-expands as it evolves toward the Hertzsprung Gap, refilling its Roche lobe and re-initiating mass transfer, i.e., Case AB mass transfer (phase 5). The donor quickly exhausts its hydrogen envelope on a thermal timescale due to rapid mass transfer to the accretor and then rapidly shrinks (phase 6), eventually contracting into a white dwarf (phase 7). 

The detached post-Algol phase is a non-negligible fraction in the lifetime of an interacting binary. The post-Algol phase occurs on a nuclear timescale along the TAMS as the hydrogen core is gradually exhausted and the subgiant slowly contracts. For the given binary evolution model of \citet[][see their Fig.~3]{Pols1994}, Case~A mass transfer lasts 4.4 Myr, the detached post-Algol phase lasts 0.2 Myr, and Case AB mass transfer lasts only 0.1 Myr. The proportions are very similar in the binary evolution models of \citet[][see their Fig.~4]{Siess2018} and \citet[][see their Fig.~2]{Sen2022}. For systems that evolve through both Case A and AB mass transfer, the in-between detached post-Algol phase lasts roughly 5\% of the semi-detached mass-transfering phases. For every 100 semi-detached Algols undergoing slow Case A mass transfer, we naturally expect to find a few detached post-Algols.

Conversely, the evolution from a cool subgiant to a hot stripped star or white dwarf (WD) occurs on a rapid thermal timescale (phase 6 in Fig.~\ref{fig:evTrack}). According to the binary evolution model of \citet[][see their Fig.~4]{Siess2018}, the subgiant donor contracts  from $RLFF$ = 100\% at the end of Case AB mass transfer to $RLFF$ = 75\% in $\approx$\,50,000 yrs, which is less than 0.2\% of the corresponding Case A mass transfer phase. We would need a sample of 500 semi-detached Algols before we expect to find a single subgiant with $RLFF$~$\approx$~75\%\,-\,95\% that is shrinking on a rapid thermal timescale toward a hot compact object. Detached EBs containing a B-type MS star and an evolved subgiant ex-donor are much more likely to be post-Algols on the TAMS than post-mass-transfer systems.

We note that not all Case A survivors evolve through Case AB mass transfer. If the donor is already heavily stripped or if the orbit has widened enough, the donor may simply become a hot stripped star or WD. Nonetheless, such evolution from a cool Roche-lobe filling subgiant at the end of Case A mass transfer to a hot compact object occurs on a rapid thermal timescale.  Just as it is unlikely to observe a subgiant that is quickly contracting on a thermal timescale immediately after Case AB mass transfer, we are unlikely to detect a subgiant that is rapidly shrinking on a thermal timescale immediately after Case A mass transfer.

\section{Object Selection from OGLE}\label{sec:Obj}

We search for additional evolved, irradiated EBs in the OGLE-III LMC survey as follows. We initially select the 100,836 stars from the OGLE-III LMC photometric maps with average brightness $\langle I \rangle$~$<$~17.5 and color $\langle V \rangle - \langle I \rangle$~$<$~0.2 \citep{Udalski2008}. Given a distance $d$~=~49.6~kpc to the LMC \citep[distance modulus $\mu$~=~18.48;][]{Pietrzynski2019}, our magnitude and color cuts correspond to OB stars. Of the selected OB stars, \citet{Gracyk2011} analyzed the 98,776 (98\%) with $N_{\rm I}$~$\ge$~120 I-band photometric measurements for variability and eclipses. They identified 3,523 EBs across orbital periods P~=~1\,-\,10~days within this photometrically selected subset. Thus 3.6\% of OB stars in the LMC are EBs with $P$ = 1\,-\,10\,days (see also \citealt{Moe2013}).

As introduced in \citet{Moe2015a,Moe2015b} and expanded in \citet{Mowlavi2017}, we fit double Gaussian analytic models to the OGLE-III EB light curves. Our analytic model contains 10 free parameters: orbital period $P$, epoch of primary eclipse $t_{\rm o}$ (Julian date JD $-$ 2450000), I-band magnitude immediately outside of primary eclipse $I_{\rm Out}$, depth of primary eclipse $\Delta I_1$, standard deviation of primary eclipse $\Theta_1$, the corresponding phase, depth, and standard deviation of the secondary eclipse $\Phi_2$, $\Delta I_2$, and $\Theta_2$, respectively, the full amplitude $\Delta I_{\rm Ell}$ of ellipsoidal variability (two crests per orbit peaking at quadratures $\phi$ = 0.25 and 0.75), and the full amplitude $\Delta I_{\rm Refl}$ of the reflection effect (one crest per orbit peaking at $\phi$ = 0.5). For a given Julian date JD of observation, the orbital phase is first computed:

\begin{equation}
\phi = \frac{({\rm JD}-t_{\rm o})~{\rm mod}~P}{P}~~.
\end{equation}

\noindent We then compute the analytic light curve model:

\begin{multline}
  I(\phi) =   I_{\rm Out} + \Delta I_1\,{\rm exp}\Big(-\frac{\phi^2}{2\Theta_1^2}\Big)
                        +\Delta I_1\,{\rm exp}\Big(-\frac{(\phi-1)^2}{2\Theta_1^2}\Big) \\
                        +\Delta I_2\,{\rm exp}\Big(-\frac{(\phi-\Phi_2)^2}{2\Theta_2^2}\Big)
                        + \frac{\Delta I_{\rm Ell}}{2}\Big({\rm cos}(4\pi\phi)-1\Big) \\
                         + \frac{\Delta I_{\rm Refl}}{2}\Big({\rm cos}(2\pi\phi)-1\Big)
\end{multline}

\noindent across 0~$\le$~$\phi$~$<$~1. For each data point, we add a 0.005~mag systematic error in quadrature with the measurement uncertainty as motivated in \citet{Moe2015a,Moe2015b}. We utilize a Levenberg-Marquardt algorithm to minimize the $\chi^2$ statistic, and we clip up to $N_{\rm c}$ $\le$ 10 outliers that exceed 5$\sigma$ for each EB.

\begin{figure*}[hp!]
    \centering
    \includegraphics[trim = 0mm 19mm 0mm 22mm, clip=true, width=\linewidth]{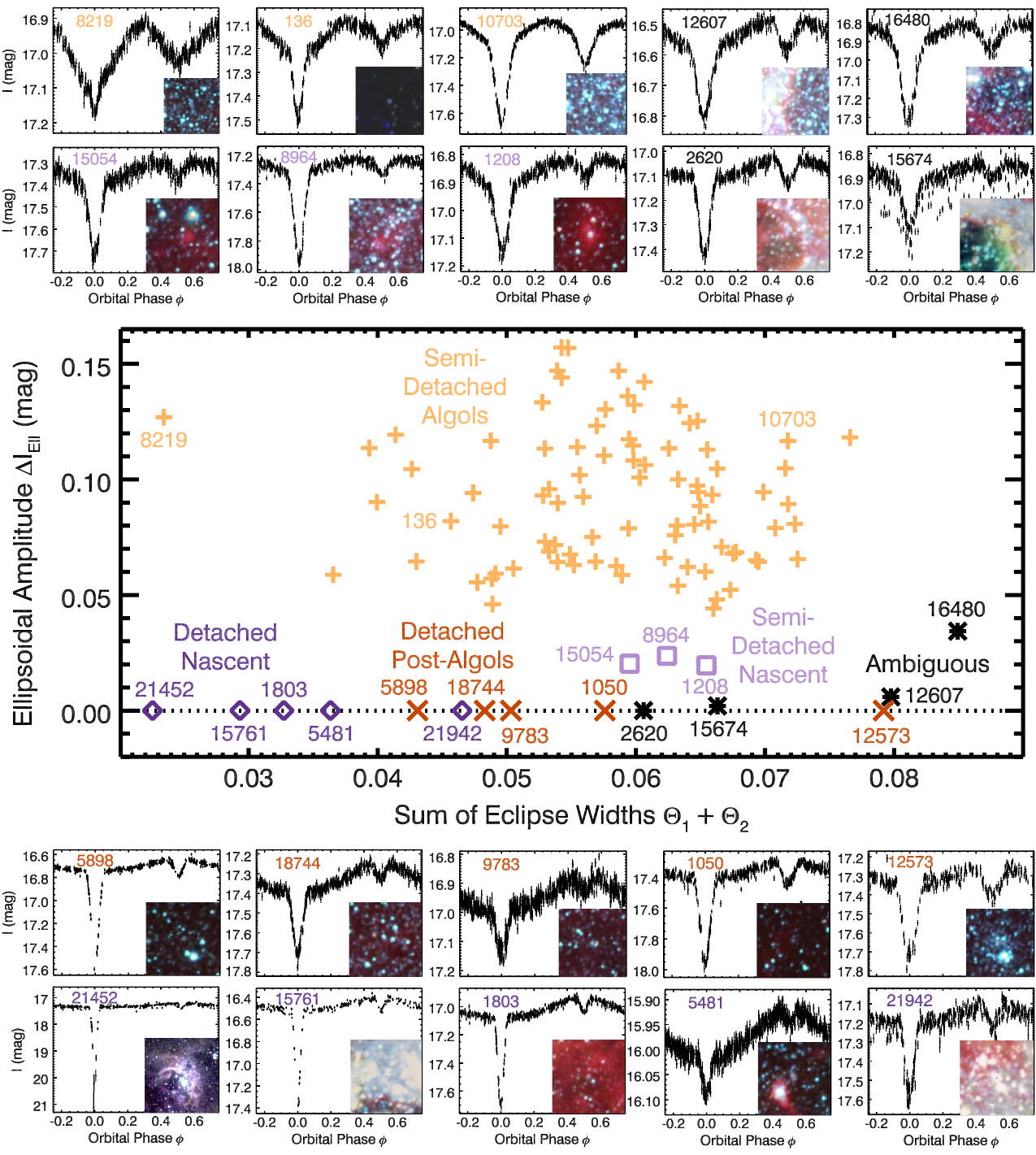}
    \caption{Ellipsoidal amplitudes versus sum of eclipse widths for the 103 OGLE-III LMC EBs with OB primaries and reflection effect amplitudes $\Delta I_{\rm Refl}$~$>$~0.08 mag. We categorize 5 detached nascent EBs (dark purple diamonds), 3 semi-detached nascent EBs (light purple squares), 86 semi-detached Algols (light orange pluses), the 5 featured detached post-Algols (red crosses), and 4 ambiguous EBs (black asterisks). We display the light curves of three Algols and all members of the other categories. For each light curve, we display a 20-pc\,$\times$\,20-pc inset from combined optical DSS and infrared Spitzer images, except for EB-21452 embedded in the Tarantula~Nebula (30~Doradus), where we show a 80-pc\,$\times$\,80-pc near-IR 2MASS image. Similar to the semi-detached Algols, our detached post-Algol candidates do not reside in star-forming environments, distinguishing them from the nascent EBs.  }
    \label{fig:SampleSelection}
\end{figure*}

We fit our 10-parameter analytic model to the OGLE-III light curves for the 3,523 selected EBs. We identify 103 EBs with large reflection effect amplitudes $\Delta I_{\rm Refl}$ $>$ 0.08 mag.  Nascent EBs with large reflection effects $\Delta I_{\rm Refl}$~$>$~0.08~mag are $\tau$~$<$~3~Myr~old and reside in bright H\,II regions \citep{Moe2015a}. Thus we can distinguish nascent EBs from post-Algols for our selected subset with $\Delta I_{\rm Refl}$~$>$~0.08~mag.

We compare their ellipsoidal amplitudes $\Delta I_{\rm Ell}$ versus their sum of eclipse widths $\Theta_1$\,+$\Theta_2$ in Fig.~\ref{fig:SampleSelection}. We plot the corresponding light curves of interesting systems labeled in the main panel. To probe the physical nature of interesting EBs, we also show the surrounding physical environments of the highlighted EBs based on images collected from Aladin Lite\footnote{https://aladin.cds.unistra.fr/AladinLite/}. For all but one EB, we display a 20-pc\,$\times$\,20-pc region by overlaying the optical DSS2 and infrared Spitzer images together. EB-21452 is deeply embedded near the center of the Tarantula Nebula (30 Doradus), and so we display a 80-pc\,$\times$\,80-pc image from the near-IR 2MASS survey for this object only.

In Fig.~\ref{fig:SampleSelection}, the semi-detached Algols (light orange pluses) comprise a distinct population with ellipsoidal amplitudes $\Delta I_{\rm Ell}$~=~0.04\,-\,0.16~mag and sum of eclipse widths $\Theta_1$\,+\,$\Theta_2$~=~0.04\,-\,0.08. Their large ellipsoidal amplitudes ($\Delta I_{\rm Ell}$ $>$ 0.04~mag) confirm that they are semi-detached. These Algols have inverted their mass ratios and are currently in the slow phase of Case~A mass transfer whereby the subgiant donors are still filling their Roche lobes. The outlier EB-8219 exhibits a small grazing primary eclipse without a secondary eclipse, resulting in the small sum $\Theta_1$\,+\,$\Theta_2$ = 0.023.  We display the light curves and environments of EB-8219 and non-grazing Algols EB-136 and EB-10703 in Fig.~\ref{fig:SampleSelection}. The significant majority of Algols, including these three, reside in old environments without detectable gas or dust emission within 10~pc.

Meanwhile, the nascent EBs are embedded in young, active star-forming regions. The detached nascent EBs (dark purple diamonds in Fig.~\ref{fig:SampleSelection}) exhibit zero ellipsoidal variability and have narrow eclipse widths $\Theta_1$\,+\,$\Theta_2$ = 0.023\,-\,0.047. \citet{Moe2015a} identified EB-1803, EB-15761, and EB-21452 as nascent EBs across $P$ = 4.0\,-\,8.2~days, while EB-5481 and EB-21942 at $P$~=~2.5~days and 2.3 days, respectively, just escaped their search criteria of $P$ $>$~3~days for well-detached systems. \citet{Moe2015a} found additional nascent EBs with smaller reflection effect amplitudes $\Delta I_{\rm Refl}$ = 0.02\,-\,0.08 mag, which tend to be slightly older. In contrast, our five detached nascent EBs with $\Delta I_{\rm Refl}$~$>$~0.08~mag are the youngest members of their class. As shown in Figs.~4~and~9 of \citet{Moe2015a}, nascent EBs with $\Delta I_{\rm Refl}$~$>$~0.08~mag are $\tau$~$<$~3~Myr~old and reside in bright, compact H\,II regions. For example, the prototype EB-1803 has the largest reflection effect amplitude of $\Delta I_{\rm Refl}$ = 0.14 mag and an estimated age of 0.7~Myr \citep{Moe2015a}. Similarly, EB-21452 is only 0.6~Myr~old and embedded near the core of the Tarantula Nebula (30 Doradus). Despite its long period of $P$ = 8.2~days, EB-21452 exhibits a substantial reflection effect of $\Delta I_{\rm Refl}$~=~0.12 mag. It also has an extremely deep primary eclipse of $\Delta I_1$ = 2.8 mag because the cool, solar-type pre-MS companion completely eclipses the hot early-B primary \citep{Moe2015a}.

\begin{deluxetable*}{lcccccccccc}[t!]
    \setlength{\tabcolsep}{4pt}
    \tablecaption{OGLE Properties of Detached Post-Algols \label{tab:OGLEproperties}}
   \tabletypesize{\footnotesize}
    \tablehead{
        \colhead{} & \colhead{} & \colhead{} & \colhead{} & \multicolumn{3}{c}{OGLE-III} & \colhead{} & \multicolumn{3}{c}{OGLE-IV} \\
        \colhead{Object} & \colhead{RA} & \colhead{Dec} & \colhead{} & \colhead{$\langle$I$\rangle$} & \colhead{$\langle$V$\rangle-\langle$I$\rangle$} & \colhead{$P$\,(days)} & \colhead{} & \colhead{$\langle$I$\rangle$} & \colhead{$\langle$V$\rangle-\langle$I$\rangle$} & \colhead{$P$\,(days)}  
    }
    \startdata
    EB-1050 & 04:48:39.69 & $-$68:27:30.9 & & 17.362 & \, \,0.012 & 1.670790 & & 17.364 & $-$0.036 & 1.6707910\\
    EB-5898 & 05:02:58.72 & $-$70:49:44.7 & & 16.704 & $-$0.053 & 5.323935 & & 16.704 & $-$0.053 & 5.3239359 \\
    EB-9783 & 05:12:06.50 & $-$67:46:22.1 & & 16.958 & $-$0.035 & 1.751504 & & 16.834 & $-$0.012 & 1.7515047 \\ 
    EB-12573 & 05:18:57.58 & $-$68:12:43.4 & & 17.307 & $-$0.026 & 1.200990 & & 17.298 & $-$0.116 & 1.2009905 \\
    EB-18744 & 05:32:39.86 & $-$68:42:56.3 & & 17.334 & $-$0.020 & 2.073370 & & 17.250 & $-$0.026 & 2.0733705\\
    \enddata
\end{deluxetable*}

We find three semi-detached nascent EBs with small but non-zero ellipsoidal amplitudes $\Delta I_{\rm Ell}$ $\approx$ 0.02~mag (light purple squares in Fig.~\ref{fig:SampleSelection}), indicating that the pre-MS companions are close to filling their Roche lobes. Compared to the detached nascent EBs, these three have wider eclipses ($\Theta_1$\,+\,$\Theta_2$ $\approx$ 0.06) and are at systematically shorter orbital periods: EB-8964 at P = 3.9 days and both EB-1208 and EB-15054 at $P$ = 1.7 days. Interestingly, the three semi-detached nascent EBs reside in compact H~II regions that are only a few pc wide (see inset images in Fig.~\ref{fig:SampleSelection}).

\begin{deluxetable*}{lcrrrrrrrrrrccc}[t!]
    \setlength{\tabcolsep}{3pt}
    \tablecaption{Analytic Fit Parameters to OGLE Light Curves of Detached Post-Algols \label{tab:OGLEfitLC}}
    \tabletypesize{\footnotesize}
    \tablehead{
        \colhead{Object} & \colhead{Survey} & \colhead{$P$} & \colhead{$t_{\rm o}$} & \colhead{$I_{\rm Out}$} &
        \colhead{$\Delta I_1$} & \colhead{$\Theta_1$} &
        \colhead{$\Phi_2$} & \colhead{$\Delta I_2$} &
        \colhead{$\Theta_2$} & \colhead{$\Delta I_{\rm Ell}$} &
        \colhead{$\Delta I_{\rm Refl}$} & \colhead{$N_{\rm I}$} &
        \colhead{$N_{\rm c}$} & \colhead{$\chi^2/\nu$} 
    }
    \startdata
    EB-1050 & OGLE-III & 1.6707865 & 3564.4381 & 17.417 & 0.542 & 0.0276 & 0.4993 & 0.140 & 0.0300 & 0.000 & 0.114 & 453 & 6 & 1.30 \\
     ~  & & $\pm$\,0.0000010 & $\pm$\,0.0006 & $\pm$\,0.003 & $\pm$\,0.007 & $\pm$\,0.0004 & $\pm$\,0.0011 & $\pm$\,0.005 & $\pm$\,0.0013 & $\pm$\,0.002 & $\pm$\,0.004 & & &  \\ 
     \cline{2-15}
       & OGLE-IV & 1.6707819 & 7001.2412 & 17.446 & 0.562 & 0.0273 & 0.5016 & 0.127 & 0.0304 & 0.000 & 0.106 & 904 & 4 & 1.92 \\ 
     ~ & & $\pm$\,0.0000003 & $\pm$\,0.0003 & $\pm$\,0.002 & $\pm$\,0.003 & $\pm$\,0.0002 & $\pm$\,0.0006 & $\pm$\,0.002 & $\pm$\,0.0007 & $\pm$\,0.002 & $\pm$\,0.002 & & & \\
    \hline 
    EB-5898  & OGLE-III & 5.3238802 & 3567.5457 & 16.747 & 0.826 & 0.0215 & 0.5004 & 0.142 & 0.0216 & 0.000 & 0.094 & 439 & 1 & 1.21 \\
      ~  & & $\pm$\,0.0000055 & $\pm$\,0.0008 & $\pm$\,0.002 & $\pm$\,0.010 & $\pm$\,0.0002 & $\pm$\,0.0007 & $\pm$\,0.005 & $\pm$\,0.0007 & $\pm$\,0.001 & $\pm$\,0.002 & & & \\
      \cline{2-15}
     & OGLE-IV & 5.3238826 & 7001.4484 & 16.798 & 0.874 & 0.0214 & 0.5010 & 0.156 & 0.0186 & 0.000 & 0.101 & 84 & 1 & 2.65 \\
     ~ & & $\pm$\,0.0000055 & $\pm$\,0.0015 & $\pm$\,0.004 & $\pm$\,0.007 & $\pm$\,0.0003 & $\pm$\,0.0011 & $\pm$\,0.006 & $\pm$\,0.0017 & $\pm$\,0.004 & $\pm$\,0.004 & & & \\
    \hline
    EB-9783 & OGLE-III & 1.7514987 & 3560.7375 & 16.998 & 0.147 & 0.0265 & 0.4985 & 0.038 & 0.0239 & 0.000 & 0.109 & 829 & 1 & 1.64 \\
      ~ & & $\pm$\,0.0000025 & $\pm$\,0.0011 & $\pm$\,0.002 & $\pm$\,0.004 & $\pm$\,0.0008 & $\pm$\,0.0023 & $\pm$\,0.003 & $\pm$\,0.0027 & $\pm$\,0.002 & $\pm$\,0.002 & & & \\
      \cline{2-15}
     & OGLE-IV & 1.7515326 & 7000.7236 & 16.938 & 0.138 & 0.0261 & 0.4983 & 0.046 & 0.0243 & 0.000 & 0.102 & 508 & 1 & 1.19 \\
    ~ & & $\pm$\,0.0000011 & $\pm$\,0.0011 & $\pm$\,0.002 & $\pm$\,0.003 & $\pm$\,0.0008 & $\pm$\,0.0019 & $\pm$\,0.004 & $\pm$\,0.0021 & $\pm$\,0.001 & $\pm$\,0.002 & & & \\
    \hline
    EB-12573 & OGLE-III & 1.2009944 & 3505.5043 & 17.351 & 0.387 & 0.0359 & 0.4995 & 0.162 & 0.0433 & 0.000 & 0.099 & 277 & 1 & 2.02 \\
      ~ & & $\pm$\,0.0000015 & $\pm$\,0.0010 & $\pm$\,0.004 & $\pm$\,0.007 & $\pm$\,0.0009 & $\pm$\,0.0018 & $\pm$\,0.007 & $\pm$\,0.0021 & $\pm$\,0.005 & $\pm$\,0.005 & & & \\
      \cline{2-15}
     & OGLE-IV & 1.2009929 & 7000.3959 & 17.371 & 0.416 & 0.0348 & 0.5002 & 0.142 & 0.0390 & 0.000 & 0.084 & 949 & 2 & 1.31 \\
    ~ & & $\pm$\,0.0000003 & $\pm$\,0.0003 & $\pm$\,0.002 & $\pm$\,0.003 & $\pm$\,0.0003 & $\pm$\,0.0008 & $\pm$\,0.003 & $\pm$\,0.0009 & $\pm$\,0.001 &$\pm$\, 0.003 & & & \\
    \hline
    EB-18744 & OGLE-III & 2.0733713 & 3563.0744 & 17.403 & 0.319 & 0.0248 & 0.4998 & 0.069 & 0.0234 & 0.000 & 0.145 & 605 & 1 & 1.40 \\
      ~ & & $\pm$\,0.0000018 & $\pm$\,0.0007 & $\pm$\,0.002 & $\pm$\,0.004 & $\pm$\,0.0004 & $\pm$\,0.0014 & $\pm$\,0.004 & $\pm$\,0.0016 & $\pm$\,0.001 & $\pm$\,0.002 & & & \\
      \cline{2-15}
     & OGLE-IV & 2.0733755 & 7000.7302 & 17.406 & 0.316 & 0.0242 & 0.5001 & 0.070 & 0.0201 & 0.000 & 0.154 & 919 & 1 & 1.38 \\
    ~ & & $\pm$\,0.0000008 & $\pm$\,0.0006 & $\pm$\,0.002 & $\pm$\,0.003 & $\pm$\,0.0003 & $\pm$\,0.0010 & $\pm$\,0.003 & $\pm$\,0.0011 & $\pm$\,0.001 & $\pm$\,0.002 \\
    \enddata
\end{deluxetable*}

We mark the five detached post-Algol EB candidates, including EB-5898, as the red crosses in Fig.~\ref{fig:SampleSelection}. We list their OGLE-III properties from \citet{Gracyk2011} in Table~\ref{tab:OGLEproperties}. In Table~\ref{tab:OGLEfitLC}, we list their analytic fit parameters based on the OGLE-III light curves. The listed uncertainties are the formal 1$\sigma$ errors from the Levenberg-Marquardt covariance matrix. We also report the goodness-of-fit statistic $\chi^2/\nu$, where the degrees of freedom $\nu$ = $N_{\rm I}-N_{\rm C}-10$ for our 10-parameter model.  The other four post-Algols have rather short orbital periods $P$ = 1.2\,-\,2.1~days compared to EB-5898 with $P$~=~5.3~days. Nonetheless, all five exhibit zero ellipsoidal modulations, indicating that the subgiant donors have recently contracted to a Roche-lobe fill factor of RLFF~=~70\%\,-\,90\%. They have tidally circularized orbits as demonstrated by the phase of secondary eclipse $\Phi$ = 0.5 and eclipse widths $\Theta_1$ $\approx$ $\Theta_2$. Similar to Algols, the five detached post-Algol candidates reside in older environments free of gas or dust emission, in stark contrast to the nascent EBs in active star-forming regions. We do not find any EBs with $\Delta I_{\rm Refl}$ $>$ 0.08~mag in old environments that have intermediate ellipsoidal amplitudes $\Delta I_{\rm Ell}$ = 0.002\,-\,0.04 mag, clearly distinguishing semi-detached Algols ($\Delta I_{\rm Ell}$ $>$\,0.04 mag) from detached post-Algols ($\Delta I_{\rm Ell}$ = 0.00 mag).

For the five detached post-Algol candidates, we fit the same analytic models to their OGLE-IV light curves \citep{Pawlak2016,Glowacki2024} and present the results in Table~\ref{tab:OGLEfitLC}. There is a small offset in the out-of-eclipse magnitudes $I_{\rm Out}$ due to the difference in photometric calibrations between the OGLE-III and OGLE-IV surveys.  For EB-9783, there appears to be a slight increase in $P$ and corresponding shift in $t_{\rm o}$ between OGLE-III and OGLE-IV. However, we fit our analytic models across sequential 2-year intervals and do not find any statistically significant trends in $P$ or $t_{\rm o}$ during the combined 23-year baseline. All other light curve parameters are consistent within $<$\,4$\sigma$ between the OGLE-III and OGLE-IV fits for all five detached post-Algol candidates. 

The lack of gas emission in the vicinity of our post-Algol candidates does not completely rule out the possibility that they are nascent EBs. Runaway OB stars can travel several tens of parsecs from their birth environments within a few Myr \citep{Gies1986,Hoogerwerf2001,Fujii2011}. In principle, our post-Algol candidates could be runaway nascent EBs that have traveled significantly from their birth environments. Nonetheless, there are other lines of evidence to suggest that at least some of our post-Algol candidates are genuinely evolved (see Section~\ref{sec:Results}). Moreover, as discussed in Section~\ref{sec:post-Algol}, given our sample size of 86 semi-detached Algols, we naturally expect to find a few detached post-Algols.

Finally, we classify four EBs in Fig.~\ref{fig:SampleSelection} as ambiguous (black asterisks). They exhibit features of both the nascent EBs and post-Algols. They have extremely short orbital periods $P$ = 1.1\,-\,1.5 days and therefore wide eclipses $\Theta_1$\,+\,$\Theta_2$ = 0.061\,-\,0.085. Three of the four display non-zero ellipsoidal variability.  EB-15674 exhibits frequent dimmings, possibly at a different period from the EB orbital period. All four of the ambiguous EBs reside in large, diffuse H~II regions. They could be nascent EBs associated with the nebulosity or evolved post-Algols that just happened to linger near long-lived star-forming environments.

\section{Spectroscopic Observations and Data Reduction} \label{sec:Obs}

We obtained multi-epoch echelle spectra of four of our identified post-Algol candidates in the LMC: EB-1050, 5898, 9783, and 18744. We observed each system 2\,-\,5 times at various orbital phases. We utilized the Magellan Inamori Kyocera Echelle (MIKE) spectrograph attached to the 6.5m Magellan-Clay Telescope at Las Campanas Observatory \citep{Bernstein2003}. The MIKE spectrograph is a high-throughput double echelle spectrograph. We reduced and analyzed only the blue side (3,800\AA\,-\,5,000\AA), which covers many hydrogen and helium absorption features of B-type stars. We used the 1.0'' slit, binned 2$\times$2, and obtained ThAr arcs for wavelength calibration at the position of each target.  We present an observation log of dates, exposure times, and seeing conditions in Table~\ref{tab:obs}. The raw 2D spectra of our post-Algol candidates featured absorption features from the B-type MS star and night-sky emission lines. Conversely, we have obtained MIKE spectra of four nascent EBs, which all exhibit nebular emission lines from the surrounding H\,II regions (Frank et al., in prep.). The absence of nebular emission lines for our four post-Algol candidates is consistent with our conclusion that they are evolved binaries.

\begin{deluxetable*}{lcccccccc}[t!]
    \tablecaption{Observation Log \label{tab:obs}}
    \tablehead{
        \colhead{Object} & \colhead{UTC Date} & \colhead{UTC Time} & \colhead{Julian Date} & \colhead{Exp. Time} & \colhead{Airmass} & \colhead{Seeing} 
        & \colhead{SNR(4225\AA)} & \colhead{SNR(4775\AA)} 
    }
    \startdata
    EB-1050 &  2017-10-25 & 06:33 & 2458051.7729 & 4\,$\times\,$1000\,s & 1.3 & 0.8$''$ & 31 & 32 \\
      & 2017-10-29 & 08:02 & 2458055.8347 & 4\,$\times$\,1000\,s & 1.3 & 0.6$''$ & 32 & 32 \\
    \hline
    EB-5898 & 2017-10-25 & 05:26 & 2458051.7264 & 3\,$\times$\,1100\,s & 1.4 & 0.8$''$ & 45 & 46 \\
      & 2017-10-26 & 05:03 & 2458052.7104 & 3\,$\times$\,1100\,s & 1.5 & 0.9$''$ & 33 & 35 \\
     & 2017-10-30 & 04:09 & 2458056.6729 & 3\,$\times$\,1100\,s & 1.6 & 0.7$''$ & 36 & 39 \\
     & 2017-10-31 & 04:29 & 2458057.6868 & 3\,$\times$\,1100\,s & 1.5 & 0.8$''$ & 46 & 49 \\
    \hline
    EB-9783 & 2017-04-02 & 01:23 & 2457845.5576 & 3\,$\times$\,1000\,s & 1.7 & 1.3$''$ & 20 & 21 \\ 
     & 2017-04-03 & 00:20 & 2457846.5139 & 3\,$\times$\,1100\,s & 1.5 & 1.1$''$ & 25 & 26 \\
     & 2017-04-04 & 02:08 & 2457847.5889 & 3\,$\times$\,1150\,s & 1.9 & 1.2$''$ & 19 & 23 \\
     & 2017-10-26 & 07:06 & 2458052.7958 & 3\,$\times$\,1200\,s & 1.3 & 0.8$''$ & 31 & 31 \\ 
     & 2017-10-29 & 04:04 & 2458055.6694 & 3\,$\times$\,1200\,s & 1.6 & 0.8$''$ & 36 & 37 \\
    \hline
    EB-18744 & 2017-10-26 & 03:25 & 2458052.6424 & 4\,$\times$\,1000\,s & 1.9 & 1.0$''$ & 20 & 22 \\
     & 2017-10-29 & 05:21 & 2458055.7229 & 4\,$\times$\,1000\,s & 1.5 & 0.7$''$ & 31 & 31 \\
     & 2017-10-30 & 05:14 & 2458056.7181 & 4\,$\times$\,1000\,s & 1.5 & 0.6$''$ & 32 & 33 \\
     & 2017-10-31 & 05:29 & 2458057.7285 & 3\,$\times$\,1000\,s & 1.4 & 0.8$''$ & 26 & 27 \\
    \enddata
\end{deluxetable*}

We reduced the data with CarPY's MIKE pipeline \citep{Kelson2000, Kelson2003}, which bias-subtracted and flat-fielded the science frames, traced the apertures, and extracted the spectra from each order. We applied heliocentric velocity corrections to the wavelength solutions for each epoch. We co-added overlapping regions from neighboring orders and normalized the spectra by fitting the continuum via a two-step process. We initially masked out the Balmer and helium absorption features and fit a fifth degree polynomial to the continuum. Small undulations in the normalized flux remained due to artifacts from combining neighboring orders. We then masked out all features below 95\% of the continuum and fit a moving boxcar to the continuum. The width of the moving boxcar was varied for each spectrum so that the empirical rms dispersion of the featureless continuum across 4,200\,-\,4,250\,\AA\ and 4,750\,-\,4800\,\AA\ matched the average reported errors from the MIKE pipeline. We list the signal-to-noise ratios (SNRs) across these two intervals for each spectrum in Table~\ref{tab:obs}. The SNRs span 19-49, where observations taken with better seeing and smaller airmass typically resulted in higher SNR.

\section{Methods} \label{sec:Methods}

\subsection{Properties of B-dwarf Primaries} \label{sec:PrimParams}

\subsubsection{Atmospheric Parameters} \label{sec:AtmParams}

In the normalized MIKE spectrum of each EB, only absorption features from the luminous B-dwarf primaries (former mass gainers) are distinguishable. We therefore fit TLUSTY BSTAR2006 atmospheric models \citep{Lanz2007} to our normalized spectra for each epoch of each object via the following.  We downloaded a grid of TLUSTY models at LMC metallicity ($\nicefrac{1}{2}$\,Z$_{\odot}$) that span effective temperatures $T_{\rm eff}$ = 15,000\,-\,30,000\,K in increments of 1,000\,K and surface gravities log\,$g$ = 3.0\,-\,4.5 in increments of 0.25 dex. We smoothed the TLUSTY models with a Gaussian kernel of varying projected rotation velocities, spanning \vsini\ = 30\,-\,300\,\ks\ in increments of 5\,\ks. We then applied shifts in radial velocity $v_{\rm r}$ in spacings of 1\,\ks. We minimized the \cs statistic between each spectrum and our 4D grid of models. For each epoch, we report in Table~\ref{tab:AtmFitResults} the reduced $\chi^2/\nu$ statistic and best-fit values for $T_{\rm eff}$, log\,$g$, \vsini, and $v_{\rm r}$.

\begin{figure*}[phtb!]
    \centering
    \includegraphics[width=\linewidth]{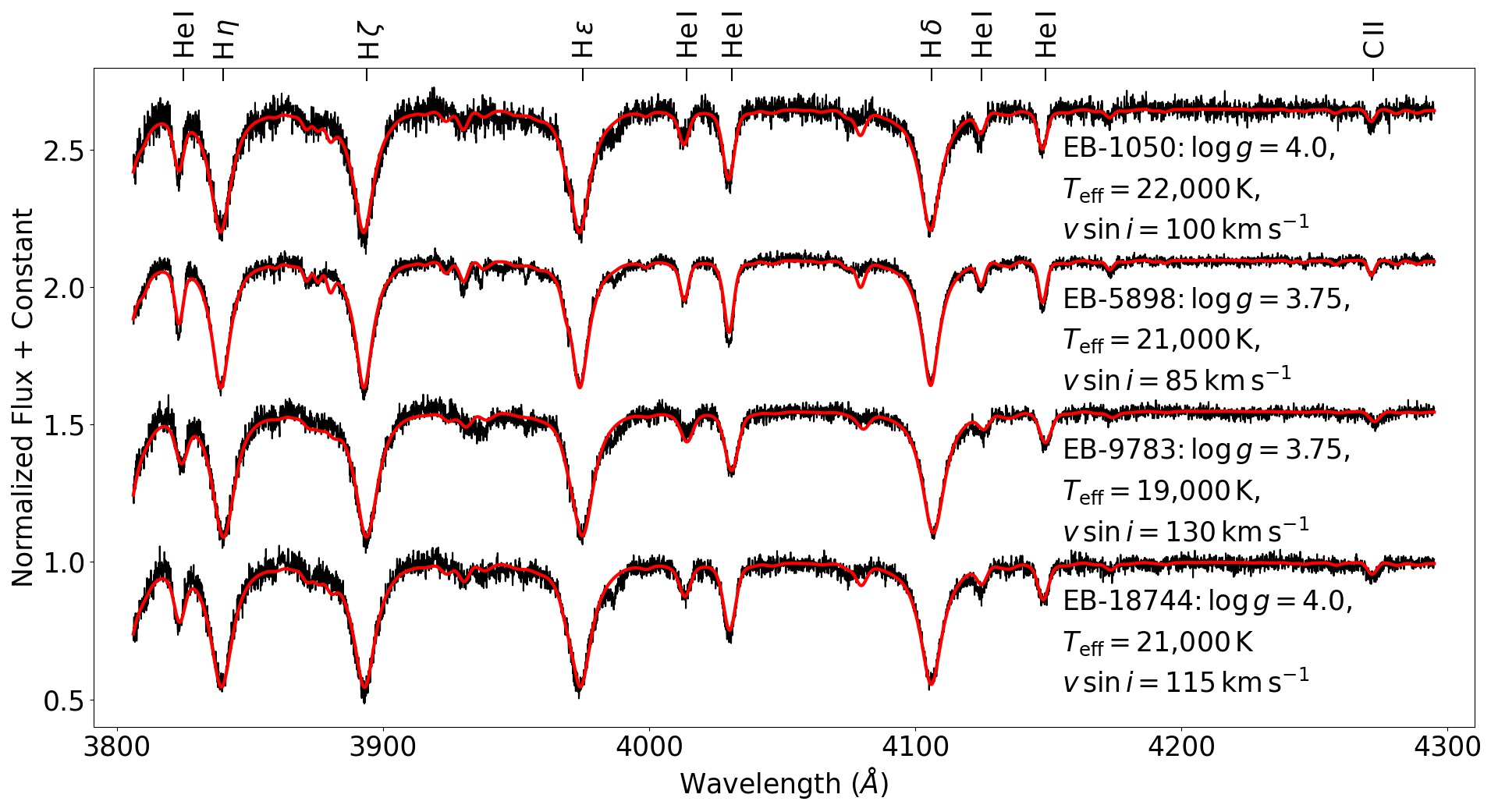}
    \caption{Stacked spectrum of each EB (black) with corresponding best-fit TLUSTY model (red) across 3800\,-\,4300\,\AA. }
    \label{fig:multiSpectra1}
\end{figure*}

\begin{figure*}[phbt!]
    \centering
    \includegraphics[width=\linewidth]{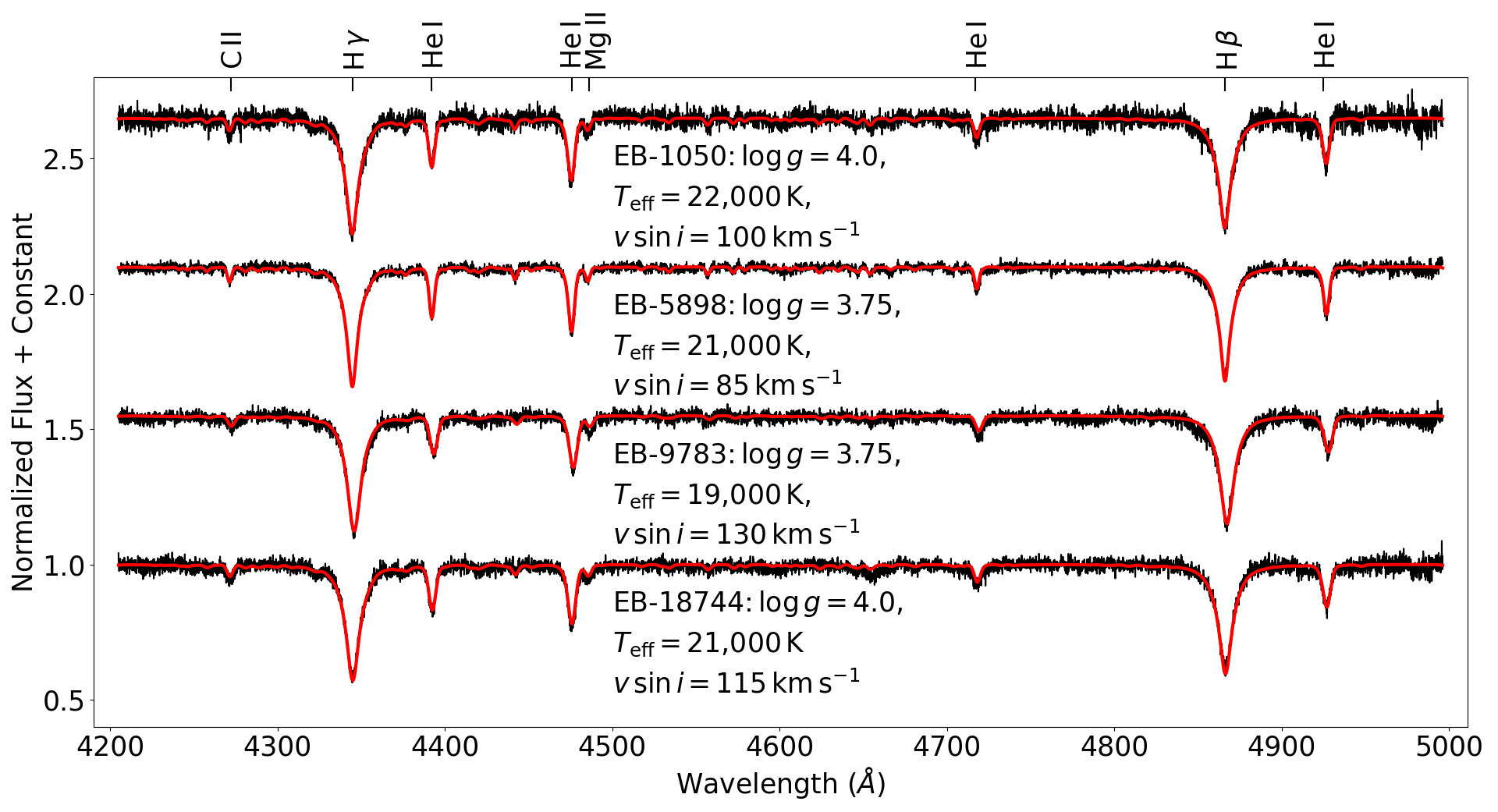}
    \caption{Same as Figure~\ref{fig:multiSpectra1} but for a wavelength range of 4200\,-\,5000\,\AA.}
    \label{fig:multiSpectra2}
\end{figure*}

For each object, we shift each spectrum by its measured radial velocity and stack the multiple epochs into a combined normalized spectrum. We also fit TLUSTY models to our combined stacked spectrum and report the best-fit parameters in Table~\ref{tab:AtmFitResults}. We plot the stacked normalized spectrum of each EB and its best-fit TLUSTY model in Figs.~\ref{fig:multiSpectra1} and \ref{fig:multiSpectra2}. The TLUSTY models successfully match the absorption profiles of both the deep Balmer hydrogen features and the shallower helium lines. All our fits result in good fit statistics with $\chi^2/\nu$ $\approx$ 1. We adopt the averages of the measured parameters across the multiple epochs. The fitted surface gravities and effective temperatures are generally consistent across all the epochs, and so we adopt overall uncertainties that are slightly smaller than the TLUSTY grid spacings. For the projected surface velocities, we add a systematic error of 3~km~s$^{-1}$ in quadrature with the standard deviation of measurements in order to estimate the final adopted uncertainty. We summarize our adopted atmospheric parameters and their uncertainties for each EB in Table~\ref{tab:AtmFitResults}. All four primaries are mid-B MS stars with $T_{\rm eff}$ = 18,000\,-\,22,000\,K and log\,$g$ = 3.75\,-\,4.00. We analyze the radial velocities in detail in Section~\ref{sec:RVcurves}.

Our B-dwarf primaries are moderate rotators, spanning \vsini\ = 86\,-\,131\,km\,s$^{-1}$, consistent with the distribution of single B-dwarfs \citep{Abt2002}. One might expect that our former B-dwarf mass accretors should be rapidly rotating, but most of the angular momentum gain occurs subsequently during Case AB mass transfer. As shown in panel g of Fig.~2 in \citet{Sen2022}, the accretor starts off synchronized and spins up slightly during fast Case A mass transfer. During slow Case A mass transfer, the accretor actually spins down and recovers synchronization by the time it becomes a detached post-Algol. Only during the subsequent Case AB mass transfer phase does the accretor spin up to a substantial fraction of break-up velocity. It is thus not surprising that the B-dwarf primaries in our post-Algol candidates are only modest rotators.

As previously indicated, the six Balmer lines and nine helium absorption features all match the profiles of our best-fit TLUSTY model. The shallow C II 4267\AA\ and Mg II 4481\AA\ lines also match the depths of our models. This consistency suggests that the B-dwarf surfaces have LMC abundances ($\nicefrac{1}{2}$\,Z$_{\odot}$). The surface helium abundance of mass gainers increases during mass transfer, but only slightly and mostly during Case AB mass transfer. For example, as shown in panel e of Fig.~2 in \citet{Sen2022}, the surface helium abundance of the accretor rises from an initial value of Y$_{\rm s}$ = 0.23 to Y$_{\rm s}$ = 0.24 at the end of Case A mass transfer, and then spikes to Y$_{\rm s}$ = 0.35 during Case AB mass transfer. The LMC abundances of our B-dwarf primaries are consistent with our interpretation that they are post-Algols.

\begin{deluxetable*}{lcccccc}[ht!]
    \setlength{\tabcolsep}{3pt}
    \tablecaption{Atmospheric Model Fits of B-dwarf Primaries \label{tab:AtmFitResults}}
    \tablehead{
        \colhead{Object} & \colhead{Julian Date} & \colhead{$T_{\rm eff, 1}$ (K)} & \colhead{log\,$g$} & \colhead{$v$\,sin\,$i$ (km\,s$^{-1}$)} &  \colhead{$v_{\rm r}$ (km\,s$^{-1}$)} & \colhead{$\chi^2/\nu$}
    }
    \startdata
    EB-1050 & 2458051.7729 & 22,000 & 4.0 & 95 & 310 & 0.951 \\
     & 2458055.8347 & 22,000 & 4.0 & 100 & 205 & 0.912 \\
     & Stacked & 22,000 & 4.0 & 100 & - & 0.842 \\
     & Adopted & 22,000\,$\pm$\,800 & 4.0\,$\pm$\,0.15 & 98\,$\pm$\,5 & - & - \\
    \hline
    EB-5898 & 2458051.7264 & 21,000 & 3.75 & 85 & 217 & 0.975 \\
     & 2458052.7104 & 21,000 & 3.75 & 85 & 244 & 1.011 \\
     & 2458056.6729 & 21,000 & 3.75 & 85 & 217 & 0.999 \\
     & 2458057.6868 & 21,000 & 3.75 & 90 & 229 & 1.045 \\
     & Stacked & 21,000 & 3.75 & 85 & - & 1.140 \\
     & Adopted & 21,000\,$\pm$\,800 & 3.75\,$\pm$\,0.15 & 86\,$\pm$\,4 & - & - \\
    \hline 
    EB-9783 & 2457845.5576 & 18,000 & 3.75 & 140 & 259 & 0.996 \\ 
     & 2457846.5139 & 18,000 & 3.75 & 125 & 319 & 1.024 \\
     & 2457847.5889 & 19,000 & 3.75 & 125 & 292 & 1.028 \\
     & 2458052.7958 & 19,000 & 3.75 & 135 & 331 & 1.029 \\ 
     & 2458055.6694 & 19,000 & 3.75 & 130 & 250 & 0.985 \\
     & Stacked      & 19,000 & 3.75 & 130 & - & 0.944 \\
     & Adopted      & 18,700\,$\pm$\,800 & 3.75\,$\pm$\,0.15 & 131\,$\pm$\,7 & - & - \\
    \hline
    EB-18744 & 2458052.6424 & 22,000 & 4.0 & 120 & 226 & 0.948 \\
     & 2458055.7229 & 21,000 & 4.0 & 110 & 331 & 0.924 \\
     & 2458056.7181 & 21,000 & 4.0 & 110 & 232  & 0.930 \\
     & 2458057.7285 & 21,000 & 4.0 & 115 & 331 & 0.903 \\
     & Stacked & 21,000 & 4.0 & 115 & - & 0.850 \\
     & Adopted & 21,200\,$\pm$\,800 & 4.0\,$\pm$\,0.15 & 114\,$\pm$\,5 & - & - \\
    \enddata
\end{deluxetable*}

\subsubsection{SED Fitting of B-dwarf Primaries}

\begin{figure*}
    \centering
    \begin{subfigure}{0.49\linewidth}
        \includegraphics[width=\linewidth]{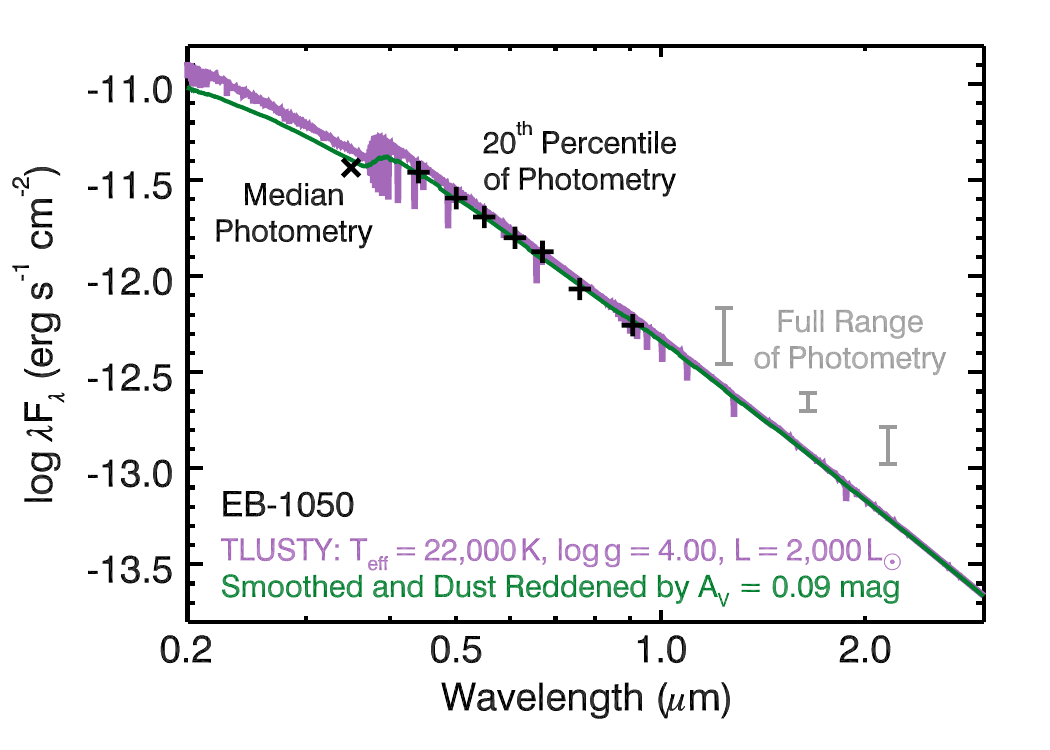}
        \caption{EB-1050}
        \label{fig:EB1050SED}
    \end{subfigure}
    \hfill
    \begin{subfigure}{0.49\linewidth}
        \includegraphics[width=\linewidth]{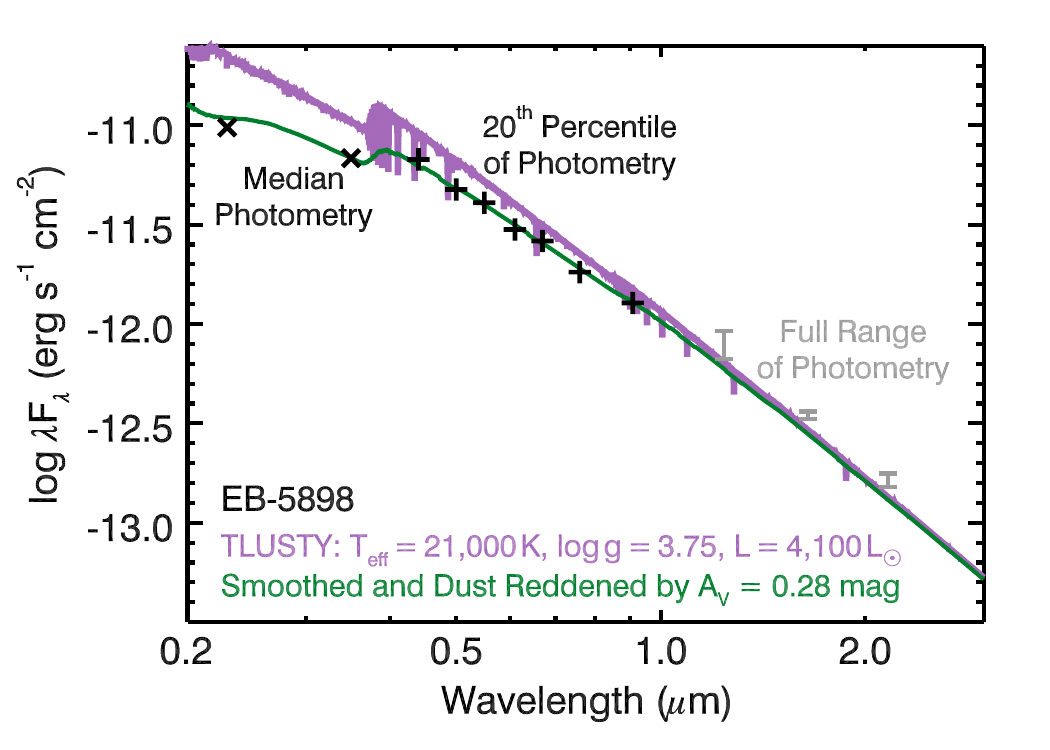}
        \caption{EB-5898}
        \label{fig:EB5898SED}
    \end{subfigure}
    \hfill
    \begin{subfigure}{0.49\linewidth}
        \includegraphics[width=\linewidth]{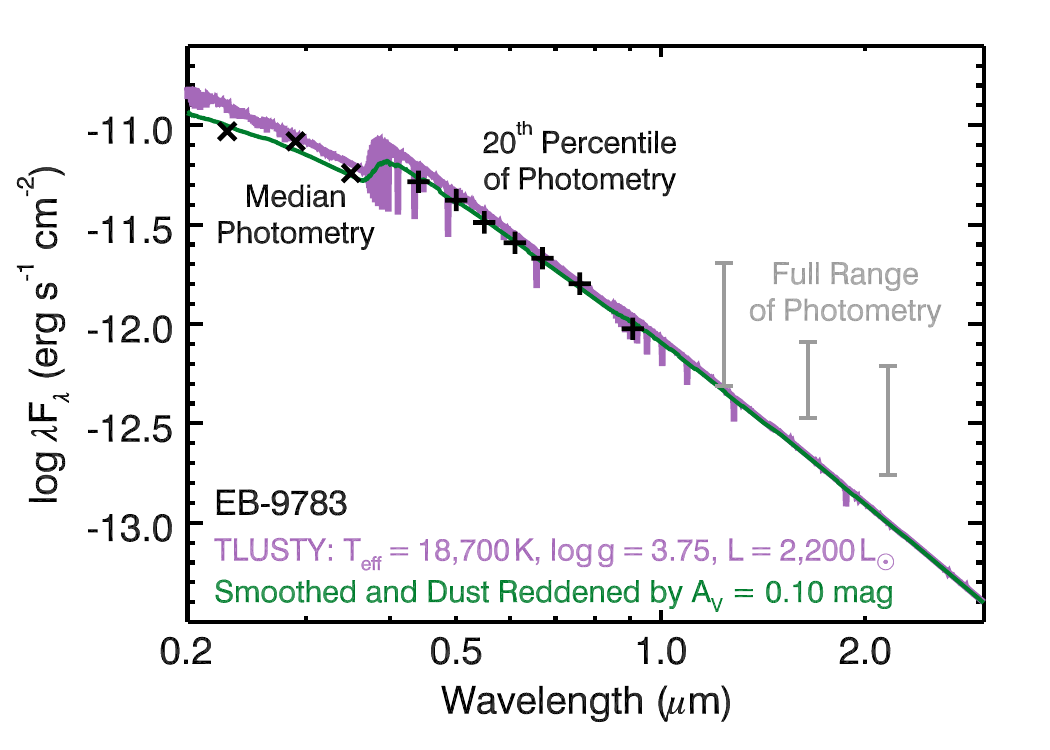}
        \caption{EB-9783}
        \label{fig:EB9783SED}
    \end{subfigure}
    \hfill
    \begin{subfigure}{0.49\linewidth}
        \includegraphics[width=\linewidth]{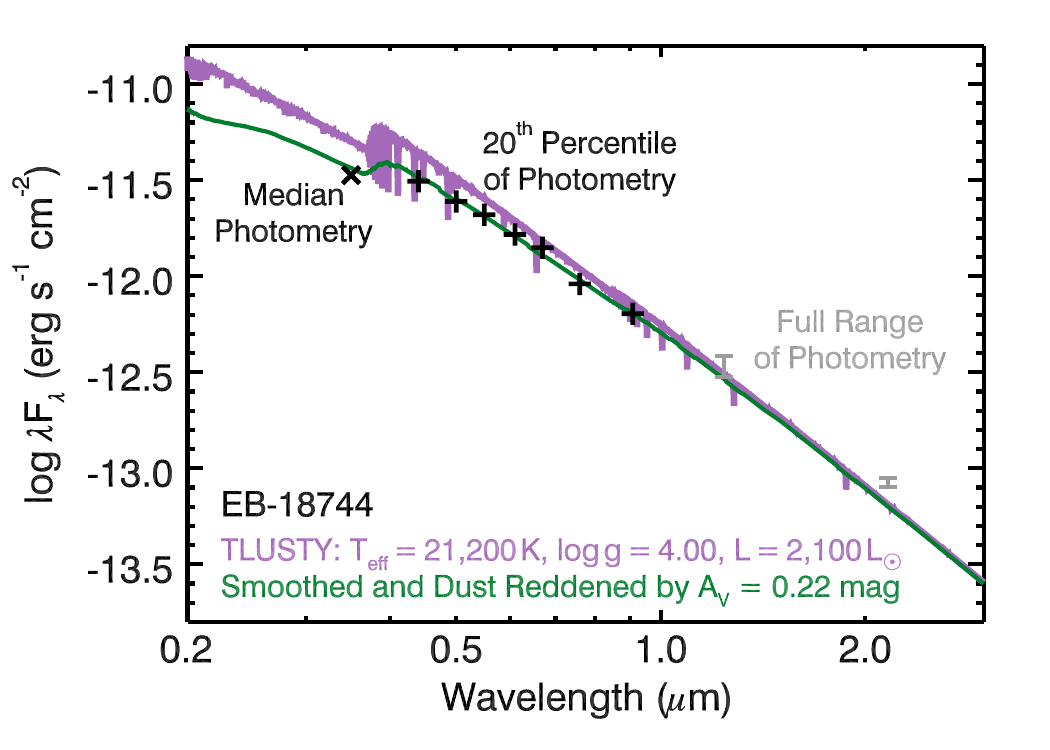}
        \caption{EB-18744}
        \label{fig:EB18744SED}
    \end{subfigure}
    \caption{TLUSTY model atmospheres (purple) and smoothed, dust-reddened SED fits (green) to UV (crosses) and optical (pluses) broadband photometry. The near-IR photometry (grey) exhibits extreme variability due to the heated side of the subgiant secondary, which we exclude when fitting the SEDs of the B-dwarf primaries.}
    \label{fig:SEDs}
\end{figure*}

We next estimate the luminosities and radii of our B-dwarf primaries by fitting SEDs to their broadband photometry. We collect all available photometry that spans the UV through near-IR from the VizieR Photometry Viewer\footnote{http://vizier.cds.unistra.fr/vizier/sed/}. Considering the variable nature of EBs and our goal of fitting the SEDs of the hot B-dwarf components alone, we select the photometry as follows. For each UV passband available, we adopt the median photometric measurements, assuming 3\% uncertainties. The well-sampled optical photometry exhibits more noticeable scatter, mostly due to phase variations of the irradiated side of the subgiant but also due to the eclipses. For optical passbands, we adopt the 20$^{\rm th}$ percentile of photometry, assuming 2\% uncertainties. The adopted percentiles roughly correspond to the phase of the EB just outside of primary eclipse, and the adopted uncertainties result in good fit statistics $\chi^2$/$\nu$ $\approx$ 1 (see below). Finally, the near-IR photometry varies significantly, depending on the orientation of the heated side of the subgiant.  We display the full range of JHK photometry when available (see grey data points in Fig.~\ref{fig:SEDs}), but we do not incorporate the near-IR data into our fits. The near-IR photometry exhibits relatively uniform scatter across the full range, as expected from the heated side of the subgiant gradually going into and out of view.

For each system, we interpolate the TLUSTY atmospheric models to match our adopted $T_{\rm eff}$ and log\,$g$ measured from their respective MIKE spectra (see Section~\ref{sec:AtmParams} and Table~\ref{tab:AtmFitResults}). We adopt a \citet{Fitzpatrick1999} dust-reddening law with $R_{\rm V}$ = 3.4 appropriate for LMC dust. Our SED model contains two free parameters: the scaled luminosity $L_1$ of the B-dwarf component and the V-band dust extinction $A_{\rm V}$ along the line of sight. We smooth the dust-reddened TLUSTY models with a Gaussian kernel of width 200 \AA, which roughly matches the widths of the photometric passbands. We minimize the $\chi^2$ statistic between data and model and estimate parameter uncertainties by propagating the spectroscopic errors in $T_{\rm eff}$ and log\,$g$. We display the best-fit TLUSTY models and smoothed SEDs in Fig.~\ref{fig:SEDs} for each of our four post-Algol EBs. In Table~\ref{tab:SEDfit}, we report the best-fit parameters $L_1$ and $A_{\rm V}$, the number of photometric data points $N_{\rm p}$, and the reduced $\chi^2/\nu$ statistic. We also list the corresponding I-band dust extinction A$_{\rm I}$ and dust reddening E(V$-$I) given the measured $A_{\rm V}$ and adopted dust-reddening law. For comparison, we report the dust reddening E(V$-$I)$_{\rm OGLE}$ at the coordinates of our EBs based on OGLE red clump stars \citep{Skowron2021}. Finally, we compute in Table~\ref{tab:SEDfit} the absolute magnitude M$_{\rm I}$ = I$_{\rm Out}$\,$-$\,$\mu$\,$-$\,A$_{\rm I}$ and the B-dwarf radius $R_1$ based on $T_{\rm eff,1}$, $L_1$, and the Stefan-Boltzmann law.

For all four of our post-Algol EBs, the measured dust reddenings are consistent with the values inferred from OGLE red clump stars \citep{Skowron2021}. The average dust extinction toward OB stars in the LMC is $\langle A_{\rm V} \rangle$ = 0.44~mag \citep{Zaritsky2004, Chen2022}. Meanwhile, our four evolved EBs span $A_{\rm V}$ = 0.09\,-\,0.28\,mag, all considerably less than average. The lower line-of-sight dust extinction provides further evidence that our post-Algol candidates are systematically older objects far removed from dusty star-forming regions.

\begin{deluxetable*}{lcccccccccc}[t!]
    \tablecaption{SED Fit Parameters of B-dwarf Primaries \label{tab:SEDfit}}
    \tablehead{
        \colhead{Object} & \colhead{A$_{\rm V}$ (mag)} & \colhead{$L_1$ (\Lsun)} & \colhead{$N_{\rm p}$} & \colhead{$\chi^2/\nu$} & \colhead{A$_{\rm I}$ (mag)} & \colhead{E(V$-$I)} &  \colhead{E(V$-$I)$_{\rm OGLE}$} & \colhead{M$_{\rm I}$ (mag)} & \colhead{$R_1$ (\Rsun)} & \colhead{M$_1$(\Msun)}
    }
    \startdata
    EB-1050 & 0.09\,$\pm$\,0.04 & 2,000\,$\pm$\,300 & 8 & 1.36 & 
              0.05\,$\pm$\,0.02 & 0.04\,$\pm$\,0.02 & 0.11\,$\pm$\,0.05 & 
              $-$1.11\,$\pm$\,0.04 & 3.1\,$\pm$\,0.4 & 7.1\,$\pm$\,0.6 \\
    \hline
    EB-5898 &  0.28\,$\pm$\,0.06 & 4,100\,$\pm$\,600 & 9 & 1.26 & 
               0.17\,$\pm$\,0.04 & 0.11\,$\pm$\,0.03 & 0.09\,$\pm$\,0.04 &
               $-$1.90\,$\pm$\,0.06 &  4.8\,$\pm$\,0.6 & 7.7\,$\pm$\,0.7 \\
    \hline
    EB-9783 &  0.10\,$\pm$\,0.03 & 2,200\,$\pm$\,300 & 10 & 1.12 & 
               0.06\,$\pm$\,0.02 & 0.04\,$\pm$\,0.02 & 0.09\,$\pm$\,0.05 &
               $-$1.54\,$\pm$\,0.04 &  4.3\,$\pm$\,0.6 & 6.3\,$\pm$\,0.6 \\
    \hline
    EB-18744 & 0.22\,$\pm$\,0.07 & 2,100\,$\pm$\,400 & 8 & 1.38 & 
               0.13\,$\pm$\,0.04 & 0.09\,$\pm$\,0.03 & 0.11\,$\pm$\,0.04 &
               $-$1.21\,$\pm$\,0.06 & 3.5\,$\pm$\,0.5 & 6.9\,$\pm$\,0.6  \\
    \enddata
\end{deluxetable*}

\subsubsection{Masses of B-dwarf Primaries} \label{sec:hrd1m1}

MS accretors in classical Algols follow a mass-luminosity relation that is consistent with their single-star counterparts \citep{Malkov2007,Yang2013,Negu2018,Wang2022}. Binary evolution models also demonstrate that as the accretor in an Algol gains mass, it quickly readjusts on a thermal timescale and increases in luminosity to maintain its MS structure \citep{Siess2018,vanRensbergen2021,Sen2022}. We therefore expect the B-type MS primaries in post-Algols to similarly follow the mass-luminosity relation of single stars.

\begin{figure*}[t!]
    \centering
    \begin{subfigure}{0.49\linewidth}
        \includegraphics[width=\linewidth]{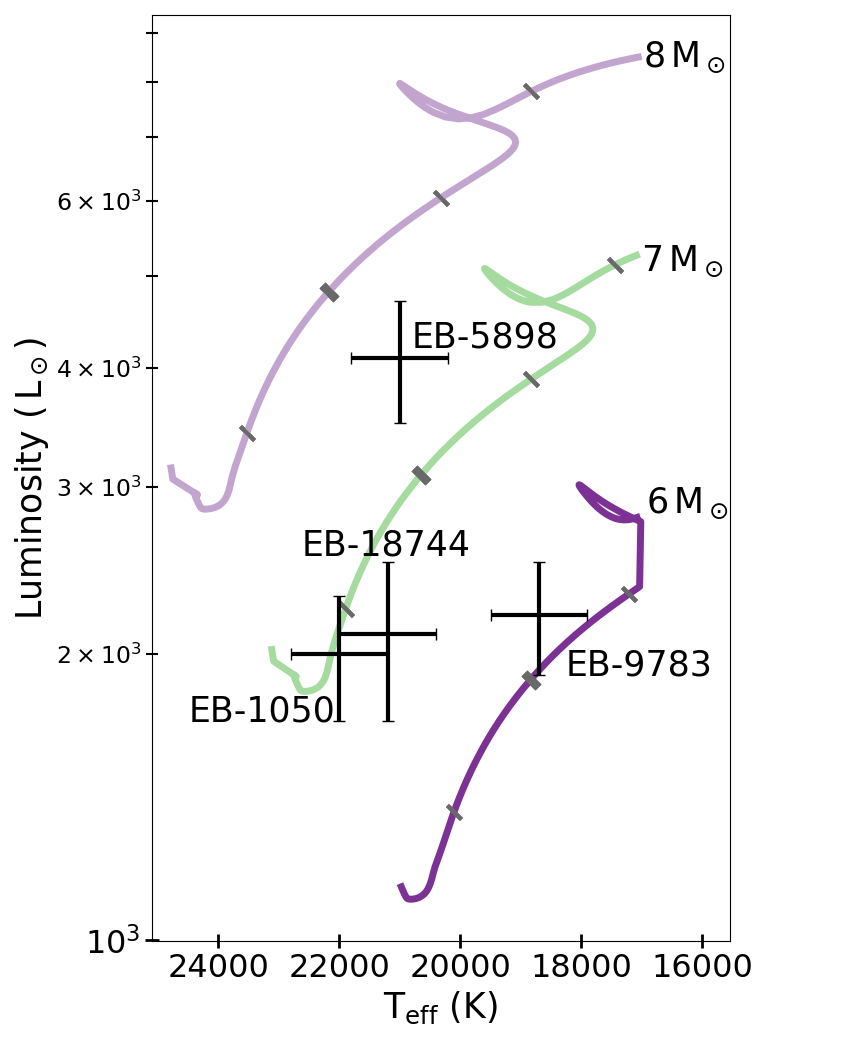}
        \label{fig:HRD1LT}
    \end{subfigure}
    \hfill
    \begin{subfigure}{0.49\linewidth}
        \includegraphics[width=\linewidth]{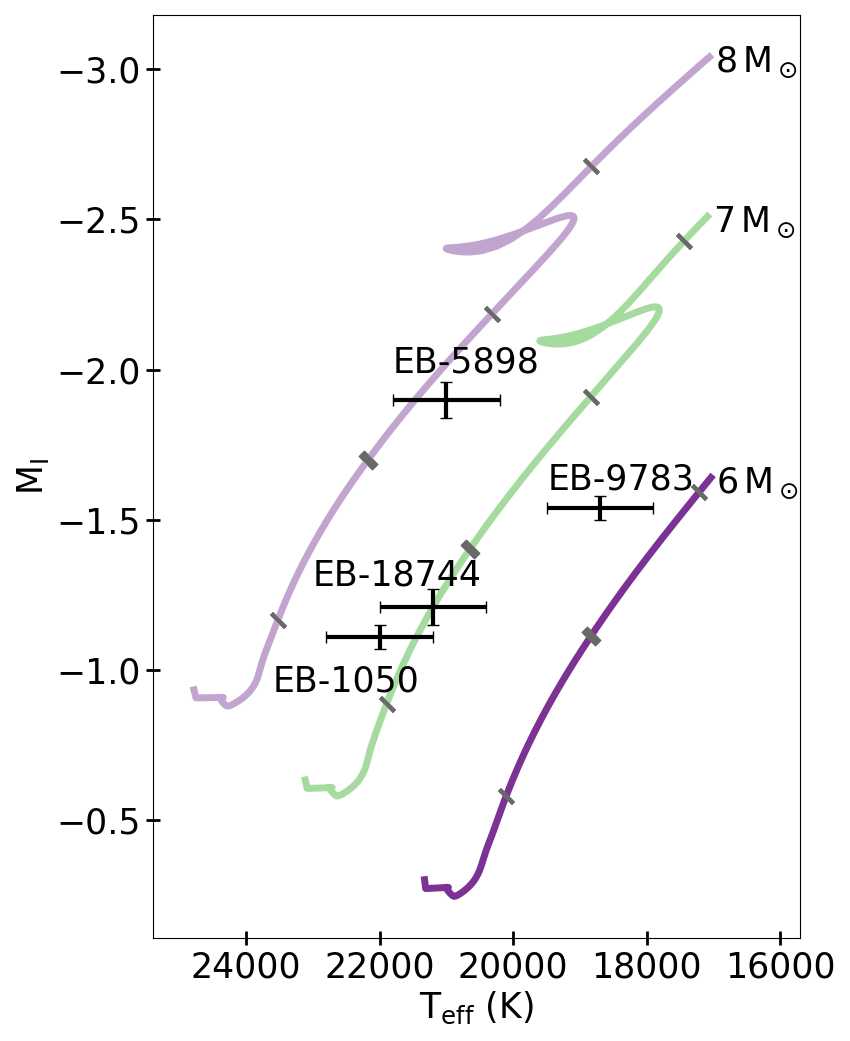}
        \label{fig:HRD1MT}
    \end{subfigure}
    \caption{Locations of our B-dwarf primaries on HR diagrams in terms of luminosity (left) and absolute magnitude (right). We overlay MIST evolutionary tracks, plotting surface gravity tick marks in intervals of 0.25 dex (thick tick mark at log\,$g$ = 4.0). }
    \label{fig:HRDPrimary}
\end{figure*}

We utilize MIST evolutionary tracks \citep{Choi2016} to estimate the masses $M_1$ of the primary B-type MS stars. In Fig.~\ref{fig:HRDPrimary}, we plot our four B-dwarf primaries on two HRDs: $L$ versus $T_{\rm eff}$ and M$_{\rm I}$ versus $T_{\rm eff}$. We overlay the MIST tracks of intermediate-mass stars at the LMC metallicity [Fe/H]\,=\,$-$0.4 and rotating at 0.4\,$v_{\rm crit}$. All four of our B-dwarf primaries are between 6 and 8\,\Msun\ and between the zero-age and terminal-age MS. The primaries in EB-5898 and EB-9783 are slightly more evolved near log\,$g$ = 3.85 according to the tracks, consistent with their measured log\,$g$ = 3.75 based on the TLUSTY atmospheric fits to their spectra. We interpolate the MIST tracks with respect to $M_1$ at the locations of our four primaries. We adopt the average $M_1$ based on the two different HRDs and report the results in the final column of Table~\ref{tab:SEDfit}. We propagate the measurement uncertainties in $L_1$, $T_{\rm eff,1}$, and M$_{\rm I}$ when deriving the measurement uncertainties in $M_1$. Considering the B-dwarf primaries are not truly evolving as single stars, we add a systematic error of 0.4\,\Msun\ in quadrature with the measurement uncertainties in order to estimate our adopted final errors in $M_1$.

\subsection{Dynamical Masses of the Subgiant Secondaries}
\label{sec:RVcurves}

\subsubsection{Radial Velocities} \label{sec:adoptedRadVel}

Given our multi-epoch MIKE spectra, we can reliably determine the dynamical masses of the subgiant secondaries in our post-Algol EBs.  We first measure radial velocities for each spectrum of each object using two different methods. As described in Section~\ref{sec:PrimParams}, we initially computed $v_{\rm r}$ by minimizing the $\chi^2$ statistic between the TLUSTY models and the entire blue region of each normalized spectrum. From our 4D $\chi^2$ grid, we marginalize the probabilities over the parameters to measure the means and uncertainties in $v_{\rm r}$ for each epoch. Second, for each normalized spectrum, we fit Gaussian profiles to two deep He\,I lines (He\,I\,4026 and He\,I\,4387) and the Doppler cores of the three  deepest Balmer lines (H\,$\beta$, H\,$\gamma$, and H\,$\delta$). We exclude shallower lines or blended features, which would have resulted in larger uncertainties. We report the best-fit $v_{\rm r}$ for each of the five lines and the overall weighted average in Table~\ref{tab:Rv}.

The radial velocities inferred from the $\chi^2$ and Gaussian fitting methods are sometimes slightly inconsistent with each other given the small 1\,-\,2 km\,s$^{-1}$ measurement errors. We therefore adopt a weighted average of the two methods, and we adopt realistic uncertainties of 3\,-\,4 km\,s$^{-1}$ that bracket both techniques. We report our adopted values and uncertainties in the final column of Table~\ref{tab:Rv}.

\begin{deluxetable*}{lccrrrrrrrr}[ht!]
    \setlength{\tabcolsep}{4pt}
    \tablecaption{Radial Velocities (km s$^{-1}$) \label{tab:Rv}}
    \tablehead{
        \colhead{Object} & \colhead{Julian} & \colhead{Phase} & \colhead{H$\delta$} & \colhead{H$\gamma$} & \colhead{H$\beta$} & \colhead{He\,I} & \colhead{He\,I} & \colhead{Weighted} & \colhead{$\chi ^2$} & \colhead{Adopted} \\
        \colhead{} & \colhead{Date} & \colhead{} & \colhead{} & \colhead{} & \colhead{} & \colhead{4026} & \colhead{4387} & \colhead{Average} & \colhead{Method} & \colhead{} 
    }
    \startdata
    EB-1050 & 2458051.7729 & 0.759 & 318.0 & 303.8 & 298.1 & 284.8 & 314.7 & 304.4 & 310.0 & 307 \\
     &  & & \,$\pm$\,4.0 & \,$\pm$\,3.2 & \,$\pm$\,4.1 & \,$\pm$\,4.7 & \,$\pm$\,4.8 & \,$\pm$\,1.8 & \,$\pm$\,1.5 & \,$\pm$\,3 \\ 
     & 2458055.8347 & 0.190 & 209.4 & 206.6 & 203.6 & 169.5 & 188.0 & 198.1 & 205.0 & 202 \\
     &  & & \,$\pm$\,4.0 & \,$\pm$\,3.4 & \,$\pm$\,4.4 & \,$\pm$\,4.7 & \,$\pm$\,4.8 & \,$\pm$\,1.9 & \,$\pm$\,1.2 & \,$\pm$\,3 \\ 
    \hline
    EB-5898 & 2458051.7264 & 0.254 & 216.7 & 214.2 & 202.2 & 201.2 & 207.7 & 208.3 & 217.0 & 215 \\
     &  & & \,$\pm$\,3.5 & \,$\pm$\,3.4 & \,$\pm$\,5.2 & \,$\pm$\,2.8 & \,$\pm$\,3.3 & \,$\pm$\,1.5 & \,$\pm$\,0.5 & \,$\pm$\,3 \\ 
     & 2458052.7104 & 0.439 & 245.2 & 238.8 & 238.2 & 220.4 & 245.7 & 236.5 & 244.0 & 240 \\
     &  & & \,$\pm$\,4.8 & \,$\pm$\,3.5 & \,$\pm$\,5.8 & \,$\pm$\,3.6 & \,$\pm$\,3.9 & \,$\pm$\,1.8 & \,$\pm$\,1.0 & \,$\pm$\,3 \\ 
     & 2458056.6729 & 0.183 & 212.7 & 212.8 & 203.0 & 203.4 & 207.9 & 208.5 & 217.0 & 213 \\
     &  & & \,$\pm$\,4.0 & \,$\pm$\,3.7 & \,$\pm$\,4.8 & \,$\pm$\,4.1 & \,$\pm$\,3.9 & \,$\pm$\,1.8 & \,$\pm$\,1.5 & \,$\pm$\,4 \\ 
     & 2458057.6868 & 0.374 & 230.7 & 229.5 & 228.3 & 213.7 & 232.5 & 226.5 & 229.0 & 228 \\
     &  & & \,$\pm$\,3.5 & \,$\pm$\,3.8 & \,$\pm$\,4.9 & \,$\pm$\,3.3 & \,$\pm$\,3.3 & \,$\pm$\,1.6 & \,$\pm$\,0.7 & \,$\pm$\,3 \\ 
    \hline
    EB-9783 & 2457845.5576 & 0.368 & 251.5 & 254.6 & 255.3 & 250.0 & 238.0 & 252.6 & 259.0 & 256 \\ 
     &  & & \,$\pm$\,7.0 & \,$\pm$\,5.4 & \,$\pm$\,6.7 & \,$\pm$\,14.6 & \,$\pm$\,12.4 & \,$\pm$\,3.4 & \,$\pm$\,2.5 & \,$\pm$\,4 \\ 
     & 2457846.5139 & 0.914 & 319.8 & 313.8 & 304.6 & 303.1 & 349.0 & 314.5 & 319.0 & 317 \\
     &  & & \,$\pm$\,4.7 & \,$\pm$\,4.4 & \,$\pm$\,4.8 & \,$\pm$\,8.1 & \,$\pm$\,8.9 & \,$\pm$\,2.4 & \,$\pm$\,1.7 & \,$\pm$\,3 \\
     & 2457847.5889 & 0.528 & 285.5 & 289.8 & 283.3 & 259.9 & 274.2 & 284.4 & 292.0 & 289 \\
     &  & & \,$\pm$\,7.7 & \,$\pm$\,4.9 & \,$\pm$\,6.1 & \,$\pm$\,13.9 & \,$\pm$\,10.9 & \,$\pm$\,3.2 & \,$\pm$\,2.4 & \,$\pm$\,4 \\
     & 2458052.7958 & 0.689 & 323.1 & 325.1 & 319.3 & 322.6 & 331.4 & 323.6 & 331.0 & 327 \\ 
     & & & \,$\pm$\,4.0 & \,$\pm$\,3.6 & \,$\pm$\,4.9 & \,$\pm$\,6.6 & \,$\pm$\,8.6 & \,$\pm$\,2.2 & \,$\pm$\,1.6 & \,$\pm$\,4 \\ 
     & 2458055.6694 & 0.329 & 243.1 & 246.7 & 248.8 & 224.6 & 233.5 & 243.1 & 250.0 & 247 \\
     & & & \,$\pm$\,3.6 & \,$\pm$\,2.9 & \,$\pm$\,3.6 & \,$\pm$\,5.6 & \,$\pm$\,5.8 & \,$\pm$\,1.7 & \,$\pm$\,1.5 & \,$\pm$\,3 \\
    \hline
    EB-18744 & 2458052.6424 & 0.350 & 213.5 & 216.0 & 229.1 & 189.0 & 231.5 & 216.9 & 226.0 & 221 \\
     &  & & \,$\pm$\,8.8 & \,$\pm$\,5.7 & \,$\pm$\,7.1 & \,$\pm$\,8.7 & \,$\pm$\,8.6 & \,$\pm$\,3.3 & \,$\pm$\,2.4 & \,$\pm$\,4 \\ 
     & 2458055.7229 & 0.835 & 332.9 & 325.3 & 323.3 & 303.1 & 326.7 & 322.7 & 331.0 & 327 \\
     &  & & \,$\pm$\,4.6 & \,$\pm$\,3.6 & \,$\pm$\,4.5 & \,$\pm$\,4.9 & \,$\pm$\,5.5 & \,$\pm$\,2.0 & \,$\pm$\,1.4 & \,$\pm$\,4 \\ 
     & 2458056.7181 & 0.315 & 229.5 & 229.2 & 226.4 & 216.6 & 219.6 & 225.1 & 232.0 & 229 \\
     &  & & \,$\pm$\,4.9 & \,$\pm$\,3.5 & \,$\pm$\,4.7 & \,$\pm$\,4.9 & \,$\pm$\,4.9 & \,$\pm$\,2.0 & \,$\pm$\,1.6 & \,$\pm$\,4 \\ 
     & 2458057.7285 & 0.803 & 328.6 & 332.6 & 326.8 & 297.0 & 320.4 & 323.8 & 331.0 & 328 \\
     &  & & \,$\pm$\,5.0 & \,$\pm$\,4.2 & \,$\pm$\,5.3 & \,$\pm$\,6.1 & \,$\pm$\,6.4 & \,$\pm$\,2.3 & \,$\pm$\,1.7 & \,$\pm$\,4 \\ 
    \enddata
\end{deluxetable*}

\subsubsection{Radial Velocity Curves} \label{sec:radVelCurves}

We next fit radial velocity curves as a function of orbital phase. We assume circular orbits based on the results of the analytic light curve fits, i.e., secondary phases are all $\Phi_2$~=~0.5 and the primary eclipse widths match the secondary eclipse widths (Section~\ref{sec:Obj}). We also adopt $P$ and $t_{\rm o}$ from the analytic fits. We therefore constrain two remaining orbital parameters: systemic velocity $\gamma$ and velocity amplitude $K_1$ of the B-type primary. We minimize the $\chi^2$ statistic between the model curves and the adopted radial velocities. We report in Table~\ref{tab:rvFits} the best-fit parameters, number of epochs $N_{\rm e}$, reduced $\chi^2/\nu$ statistic, and the probability to exceed (PTE).  Although the goodness-of-fit statistics $\chi^2/\nu$ are discrepant with unity, the PTEs are all reasonable given the small degrees of freedom $\nu$.We plot the radial velocity curves in Fig.~\ref{fig:EB1050rvlc5} (EB-1050), Fig.~\ref{fig:EB5898rvlc5} (EB-5898), Fig.~\ref{fig:EB9783rvlc5} (EB-9783), and Fig.~\ref{fig:EB18744rvlc5} (EB-18744).

The systemic velocities span $\gamma$ = 253\,-\,288\,km\,s$^{-1}$, similar to LMC's redshift of 278\,km\,s$^{-1}$ \citep{Richter1987}. The velocity semi-amplitudes span $K_1$ = 35\,-\,57 km\,s$^{-1}$, an order of magnitude larger than the corresponding velocity errors. Given the measured $K_1$, $P$, and $M_1$ and the adopted $e$ = 0, we compute $M_2$\,sin\,$i$ via the standard relation:

\begin{equation}\label{eqn:k1}
    K_1 = \left(\frac{2\pi{\rm G}}{P}\right)^{\frac{1}{3}} 
    \frac{M_2}{M_1}
    \left(M_1+M_2\right)^{\frac{1}{3}}\frac{{\rm sin}\,i}{\sqrt{1-e^2}},
\end{equation}

\noindent and we report the results in the final column of Table~\ref{tab:rvFits}. The secondary masses span $M_2$ $\approx$ $M_2$\,sin\,$i$ = 0.9\,-\,1.2\,\Msun. Our systems with $M_1$~$\approx$~7\,\Msun\ and $M_2$~$\approx$~1.0\,\Msun\ are slightly more massive than Algol, which has $M_1$ = 3.2\,\Msun\ and $M_2$ = 0.7\,\Msun\ \citep{Baron2012}.

\begin{deluxetable}{lcccccc}[h!]
    \setlength{\tabcolsep}{2pt}
    \tablecaption{Radial Velocity Fit Parameters \label{tab:rvFits}}
    \tablehead{
        \colhead{Object} & \colhead{$\gamma$ (km\,s$^{-1}$)} & \colhead{$K_\mathrm{1}$ (km\,s$^{-1}$)} & \colhead{$N_{\rm e}$} & \colhead{$\chi^2/\nu$} & \colhead{PTE} & \colhead{$M_2$\,sin\,$i$ (\Msun)}
    }
    \startdata
    EB-1050 & 253\,$\pm$\,2  & 54\,$\pm$\,2 & 2 & - & - & 1.07\,$\pm$\,0.08 \\
    \hline
    EB-5898 & 249\,$\pm$\,4  & 35\,$\pm$\,5 & 4 & 0.19 & 0.83 & 1.07\,$\pm$\,0.15 \\
    \hline
    EB-9783 & 288\,$\pm$\,2  & 47\,$\pm$\,2 & 5 & 3.61 & 0.01 & 0.86\,$\pm$\,0.07 \\ 
    \hline
    EB-18744 & 275\,$\pm$\,2  & 57\,$\pm$\,2 & 4 & 3.25 & 0.04 & 1.18\,$\pm$\,0.09 \\
    \enddata
\end{deluxetable}

\subsection{Light Curve Fitting}
\label{sec:LCfitting}

\subsubsection{PHOEBE Models}

We next fit physical light curve models by utilizing the PHysics Of Eclipsing BinariEs (PHOEBE) modeling software\footnote{https://phoebe-project.org/} \citep{Prsa2005}. We synthesize I-band light curves with PHOEBE at the times/phases of the OGLE data. We fit the OGLE-III and OGLE-IV I-band light curves separately.

For each model, we adopt $P$ and $t_{\rm o}$ determined from the analytic fits (Table~\ref{tab:OGLEfitLC}), primary effective temperature $T_{\rm eff,1}$ based on the TLUSTY atmospheric fits to the spectra (Table~\ref{tab:AtmFitResults}), primary masses $M_1$ inferred from their locations on the HRDs (Table~\ref{tab:SEDfit}), and $M_2$\,sin\,$i$ measured from the radial velocity curves (Table~\ref{tab:rvFits}). The primary radii estimated from the SEDs are relatively uncertain ($\delta R_1$/$R_1$ $\approx$ 13\%), whereas PHOEBE can generally measure radii to an accuracy of a few percent. We thus keep $R_1$ as a free parameter in our light curve models. We set the eccentricities to zero because our post-Algols are tidally circularized (see Section~\ref{sec:Obj}).  We adopt a logarithmic limb-darkening law for both components. For the B-type primary, we interpolate the limb-darkening coefficients from Table~1 of \citet{VanHamme1993} with respect to log\,$g$ and $T_{\rm eff}$. For the cool subgiant secondary, we adopt the limb-darkening coefficients based on model atmospheres internal to PHOEBE. Finally, we set the primary albedo to unity, i.e., $A_1$ = 1.0, which is appropriate for the hot radiative atmospheres of our mid-B primaries. Our PHOEBE models contain five free parameters: primary B-dwarf radius $R_1$, secondary subgiant radius $R_2$, secondary effective temperature $T_{\rm eff,2}$, secondary albedo $A_2$, and inclination $i$ of the EB. To compute each PHOEBE model, we utilize the MedicineBow cluster from the University of Wyoming's Advanced Research Computing Center (ARCC) \citep{UWARCC}.

We minimize the $\chi^2$ statistic between OGLE data and PHOEBE models via a three-step process. First, we create a dense grid of PHOEBE light curve models within our 5D parameter space. Typical spacings are 0.05\,\Rsun\ in $R_1$, 0.025\,\Rsun\ in $R_2$, 200\,K in $T_{\rm eff,2}$, 5\% in $A_2$, and 0.1$^{\circ}$ in $i$. The grids are sufficiently dense such that the best-fit model is rather close to the global solution. We marginalize over the grid to measure the probability density functions (PDFs) of each of our five parameters. We compute the weighted average for each parameter based on the PDFs. Second, we apply a Trust Region Reflective (TRF) algorithm via SciPy's package scipy.optimize.least\_squares \citep{SciPy}. We start the TRF routine with our weighted-average solutions measured from the PDFs of the dense grid, but set a lower bound in the secondary temperature to $T_{\rm eff,2}$ = 3,500 K. Finally, we utilize a Levenberg-Marquardt (LM) technique to minimize the $\chi^2$ statistic via the scipy.optimize.least\_squares package. We start the LM chain with our solution from the TRF algorithm, but now allow all free parameters to converge to any final value. Similar to our analytic models, we clip outliers in the light curve that exceed 5$\sigma$ from our best-fit PHOEBE model. We apply this three-step process for the OGLE-III and OGLE-IV light curves separately.

\subsubsection{Results of Light Curve Fits}\label{sec:ResultsLCFit}

\begin{figure*}[pt!]
    \centering
    \includegraphics[width=\linewidth]{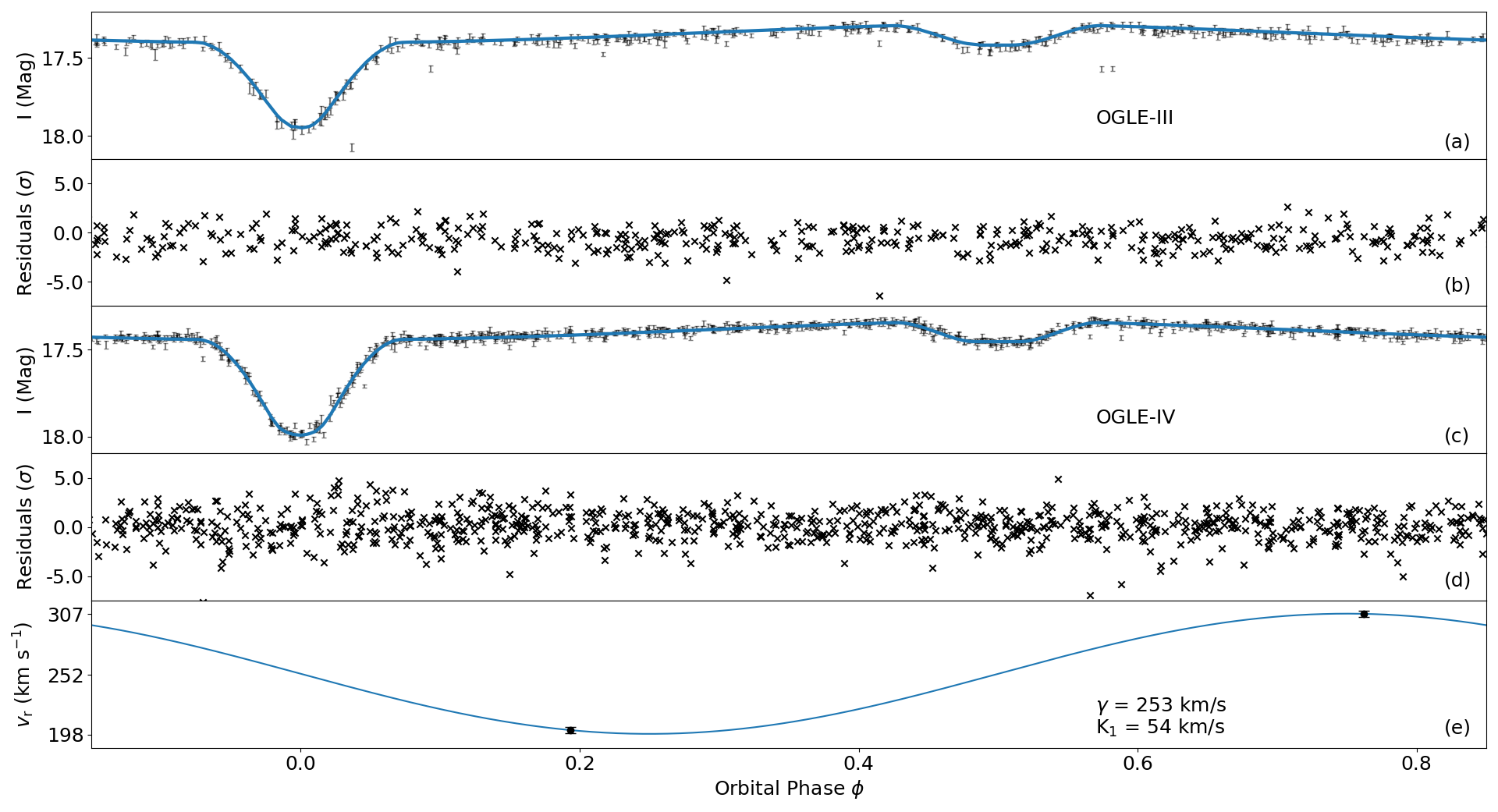}
    \caption{EB-1050 light curves and radial velocity fits. Panel \textit{a} shows the light curve from OGLE-III (black) and the corresponding PHOEBE fit (blue). Panel \textit{b} shows the residuals between the OGLE-III light curve and PHOEBE fit. Panels \textit{c} and \textit{d} show the same as Panels \textit{a} and \textit{b} but for the OGLE-IV light curve. Panel \textit{e} shows the adopted radial velocities (black) and corresponding fit (blue). }
    \label{fig:EB1050rvlc5}
\end{figure*}

\begin{figure*}[pb!]
    \centering
    \includegraphics[width=\linewidth]{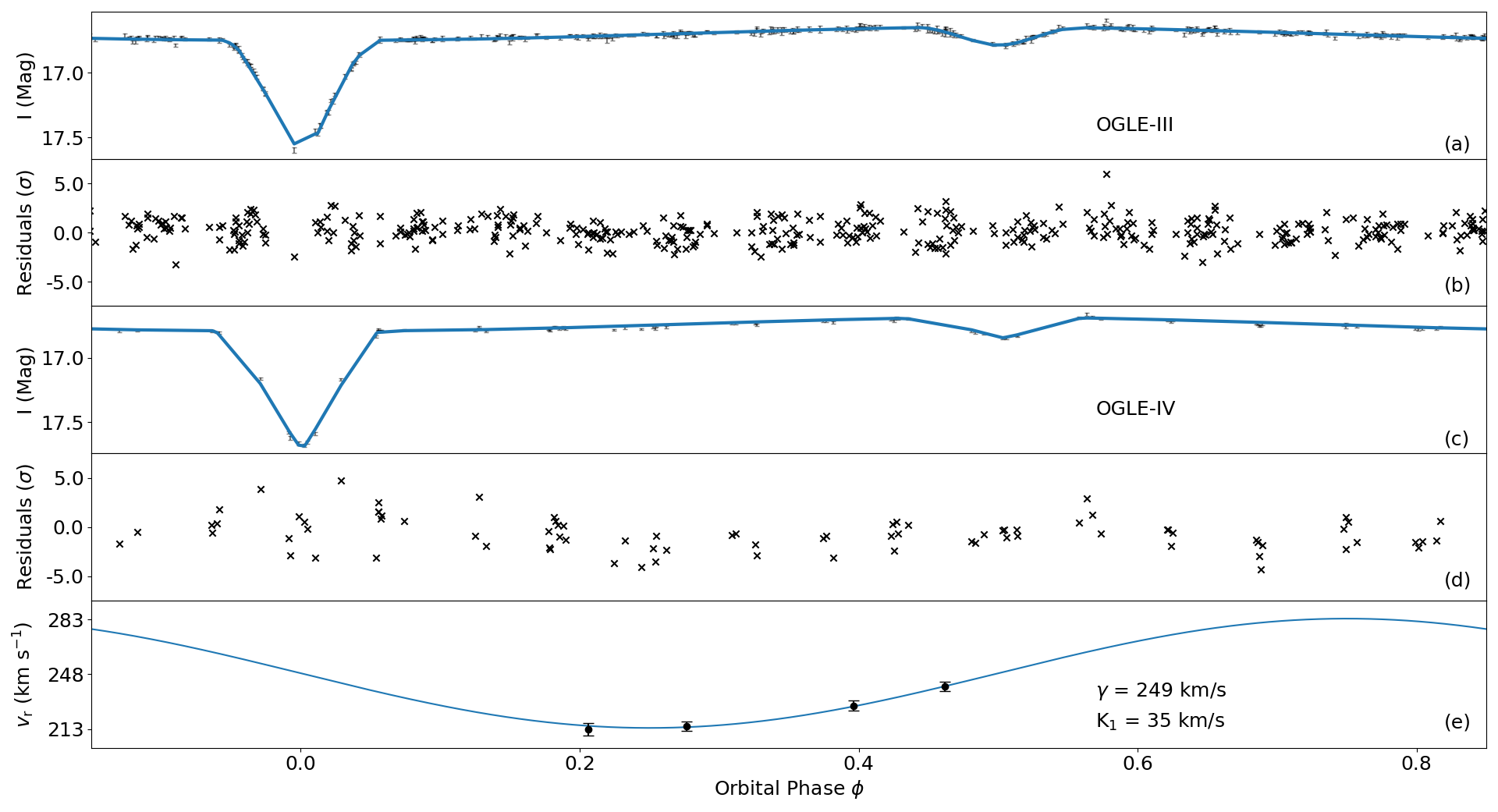}
    \caption{Same as Figure~\ref{fig:EB1050rvlc5} but for EB-5898}
    \label{fig:EB5898rvlc5}
\end{figure*}

\begin{figure*}[pt!]
    \centering
    \includegraphics[width=\linewidth]{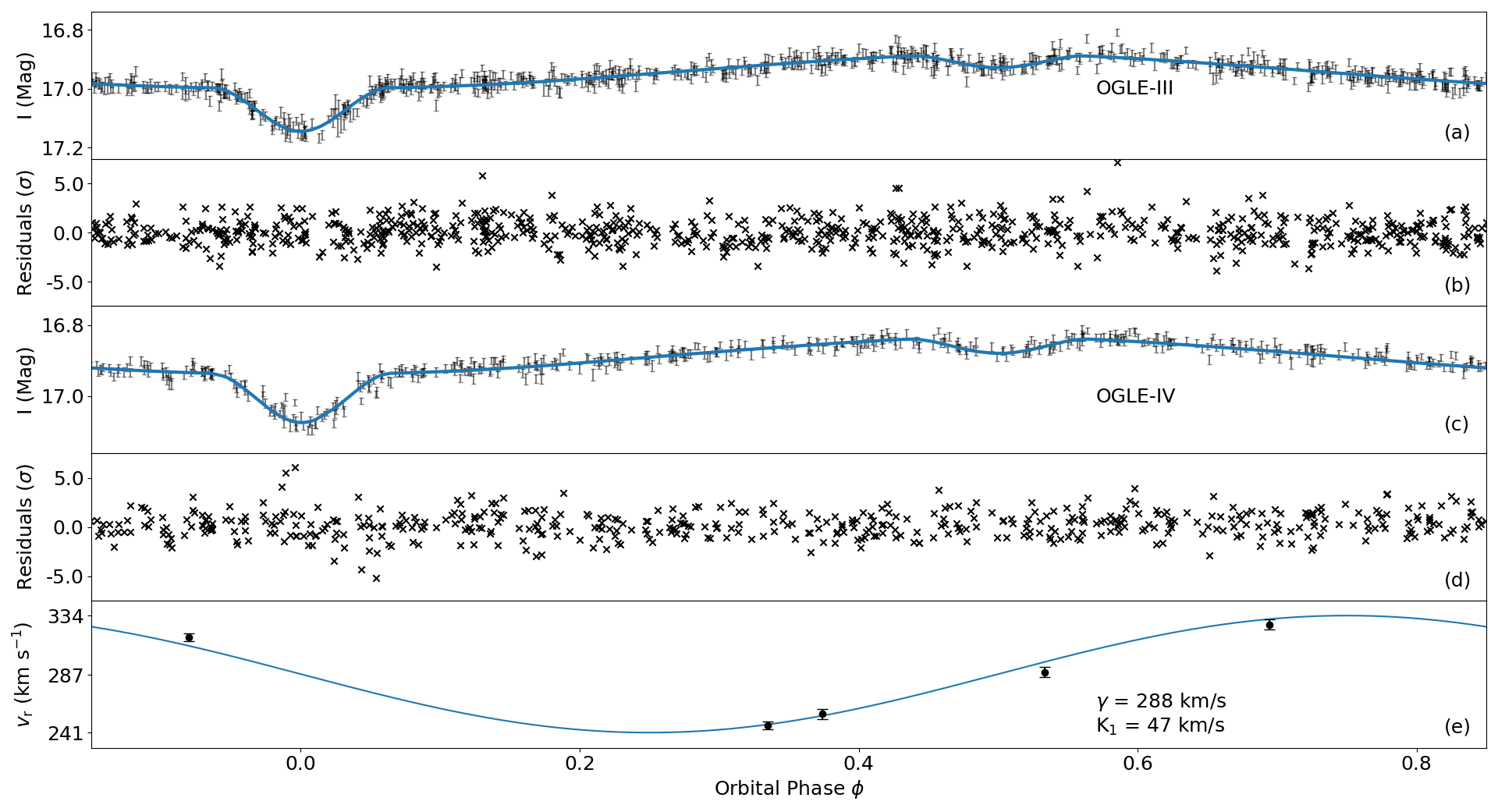}
    \caption{Same as Figure~\ref{fig:EB1050rvlc5} but for EB-9783}
    \label{fig:EB9783rvlc5}
\end{figure*}

\begin{figure*}[pb!]
    \centering
    \includegraphics[width=\linewidth]{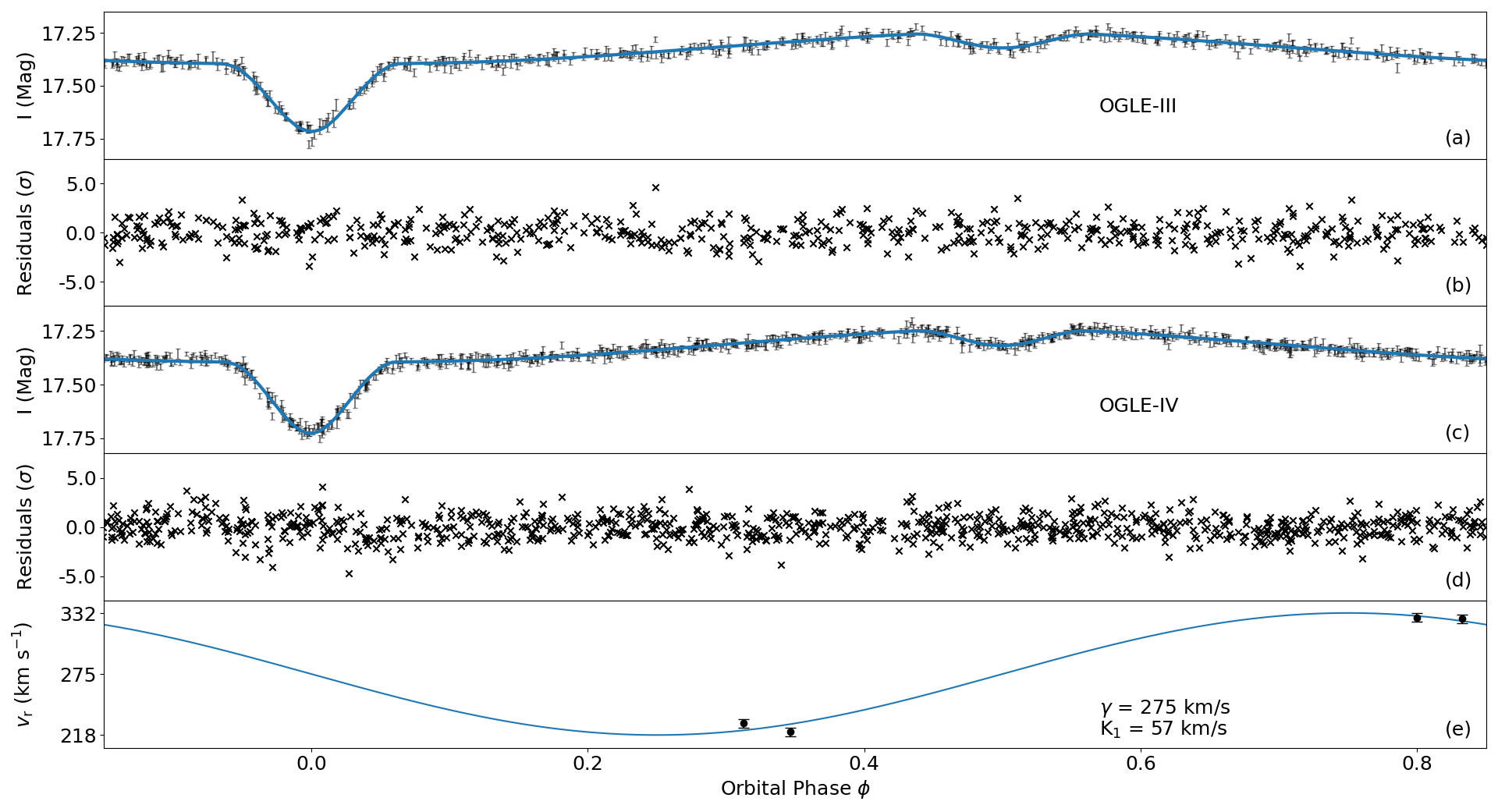}
    \caption{Same as Figure~\ref{fig:EB1050rvlc5} but for EB-18744}
    \label{fig:EB18744rvlc5}
\end{figure*}

For both OGLE-III and OGLE-IV, we report the best-fit parameters $R_1$, $R_2$, $T_{\rm eff,2}$, $A_2$, and $i$ alongside $N_{\rm I}$, $N_{\rm c}$, and $\chi^2/\nu$ in Table~\ref{tab:LMFitResults}. We also list the best-fit secondary masses $M_2$ given the measured $M_2$\,sin\,$i$ and best-fit inclinations $i$. We compute the Roche-lobe fill factors $RLFF$ for both the primaries and secondaries according to Eqn.~2 in  \citet{Eggleton1983}. We report the $RLFF$ values in the final two columns of Table ~\ref{tab:LMFitResults}. We plot the best-fit PHOEBE light curve models alongside the OGLE data and corresponding residuals in Figs.~\ref{fig:EB1050rvlc5}, ~\ref{fig:EB5898rvlc5}, ~\ref{fig:EB9783rvlc5}, ~\ref{fig:EB18744rvlc5}.

\begin{deluxetable*}{lcrrrrrrrrrrr}[hpt!]
    \tablecaption{PHOEBE light curve fits \label{tab:LMFitResults}}
    \tablehead{
        \colhead{Object} & \colhead{Survey} & \colhead{$R_\mathrm{1}$ (\Rsun)} & \colhead{$R_\mathrm{2}$ (\Rsun)} & \colhead{$T_\mathrm{eff, 2}$ (K)} & \colhead{$i$ (\textdegree)} & \colhead{A$_{\mathrm{2}}$(\%)} & \colhead{$N_{\mathrm{I}}$} & \colhead{$N_{\mathrm{C}}$} & \colhead{\cs\,/\,$\nu$} & \colhead{$M_\mathrm{2}$ (\Msun)} & \colhead{$RLFF_{\mathrm{1}}$} & \colhead{$RLFF_{\mathrm{2}}$}
    }
    \startdata
    EB-1050 & OGLE-III & 3.39 & 2.05 & 4650 & 84.1 & 62 & 453 & 6 & 1.10 & 1.08 & 0.52 & 0.74 \\
     & OGLE-IV & 3.31 & 1.95 & 4330 & 86.3 & 55 & 904 & 4 & 1.68 & 1.07 & 0.51 & 0.71 \\
     \cline{2-13}
      & Adopted & 3.37 & 2.03 & 4550 & 84.8 & 60 & - & - & - & 1.08 & 0.52 & 0.73 \\
     &  & $\pm$\,0.10 & $\pm$\,0.07 & $\pm$\,180 & $\pm$\,0.9 & $\pm$\,4  & - & - & - & $\pm$\,0.09 & $\pm$\,0.03 & $\pm$\,0.03 \\
    \hline
    EB-5898 & OGLE-III & 4.87 & 5.00 & 6080 & 83.0 & 44 & 439 & 1 & 1.00 & 1.08 & 0.33 & 0.83 \\
     & OGLE-IV & 4.88 & 4.90 & 6300 & 83.7 & 52 & 84 & 1 & 1.93 & 1.08 & 0.33 & 0.82 \\
     \cline{2-13}
      & Adopted & 4.87 & 4.98 & 6150 & 83.1 & 45 & - & - & - & 1.08 & 0.33 & 0.83 \\
     &  & $\pm$\,0.16 & $\pm$\,0.17 & $\pm$\,240 & $\pm$\,0.6 & $\pm$\,5 & - & - & - & $\pm$\,0.16 & $\pm$\,0.03 & $\pm$\,0.03 \\
    \hline
    EB-9783 & OGLE-III & 3.35 & 2.41 & 3510 & 70.9 & 47 & 829 & 2 & 1.40 & 0.91 & 0.52 & 0.89 \\
     & OGLE-IV & 3.47 & 2.40 & 4230 & 70.4 & 45 & 508 & 0 & 1.19 & 0.91 & 0.53 & 0.89 \\
     \cline{2-13}
      & Adopted & 3.43 & 2.40 & 4000 & 70.5 & 46 & - & - & - & 0.91 & 0.52 & 0.89 \\
     &  & $\pm$\,0.10 & $\pm$\,0.08 & $\pm$\,300 & $\pm$\,0.6 & $\pm$\,4 & - & - & - & $\pm$\,0.08 & $\pm$\,0.03 & $\pm$\,0.03 \\
    \hline
    EB-18744 & OGLE-III & 3.36 & 2.89 & 3630 & 75.3 & 43 & 605 & 0 & 1.17 & 1.22 & 0.46 & 0.87 \\
     & OGLE-IV & 3.20 & 2.80 & 3960 & 76.0 & 48 & 919 & 0 & 1.16 & 1.22 & 0.44 & 0.84 \\
    \cline{2-13}
      & Adopted & 3.28 & 2.84 & 3800 & 75.7 & 46 & - & - & - & 1.22 & 0.45 & 0.85 \\
     &  & $\pm$\,0.10 & $\pm$\,0.09 & $\pm$\,170 & $\pm$\,0.6 & $\pm$\,4 & - & - & - & $\pm$\,0.10 & $\pm$\,0.03 & $\pm$\,0.03 \\
    \enddata
\end{deluxetable*}

Most of our PHOEBE light curve fits result in good fit statistics $\chi^2/\nu$ $\approx$ 1 and no visible trends in the residuals. The notable exception is the fit to the OGLE-IV light curve of EB-5898, which has $\chi^2/\nu$ = 1.9. The OGLE-IV light curve of EB-5898 is sparsely sampled with only 84 data points, making fitting difficult. Fortunately, the fit to the OGLE-III light curve of EB-5898 is exceptional with $\chi^2/\nu$ = 1.00.

In Table~\ref{tab:LMFitResults}, we adopt parameters between the OGLE-III and OGLE-IV values, giving greater weight to the fit with smaller $\chi^2/\nu$.
To estimate overall uncertainties, we add the errors from the Levenberg-Marquardt covariance matrix in quadrature with half the differences between the OGLE-III and OGLE-IV fits. We also propagate the uncertainties in correlated parameters. For example, PHOEBE robustly measures the ratio $T_{\rm eff,2}$/$T_{\rm eff,1}$, and therefore we propagate the adopted uncertainty in $T_{\rm eff,1}$ when estimating the error in $T_{\rm eff,2}$. Similarly, PHOEBE precisely measures the fractional radii $R_1$/$a$ and $R_2$/$a$, and thus the uncertainties in the actual radii $R_1$ and $R_2$ depend on the uncertainty in the total mass $M_1$+$M_2$ via Kepler's third law. 

For three of our objects (EB-1050, 5898, 18744), the primary radii measured with PHOEBE are consistent with the values inferred from SED fitting within the 1$\sigma$ uncertainties. For EB-9783, the primary radii are slightly different between the two methods, but only discrepant at the 1.4$\sigma$ level. For all four EBs, the PHOEBE light curve fits result in smaller uncertainties in the primary radii. 

Cool stars with deep convective envelopes have bolometric albedos of 40\%\,-\,50\% \citep{Rucinski1969}. The measured albedos of our cool secondary subgiants span 45\%\,-\,60\%, consistent with expectations. The consistency supports that PHOEBE accurately models the irradiation and reflection effects in our systems.

\section{Discussion}
\label{sec:Results}

\begin{figure}[h!]
    \centering
    \includegraphics[width=\linewidth]{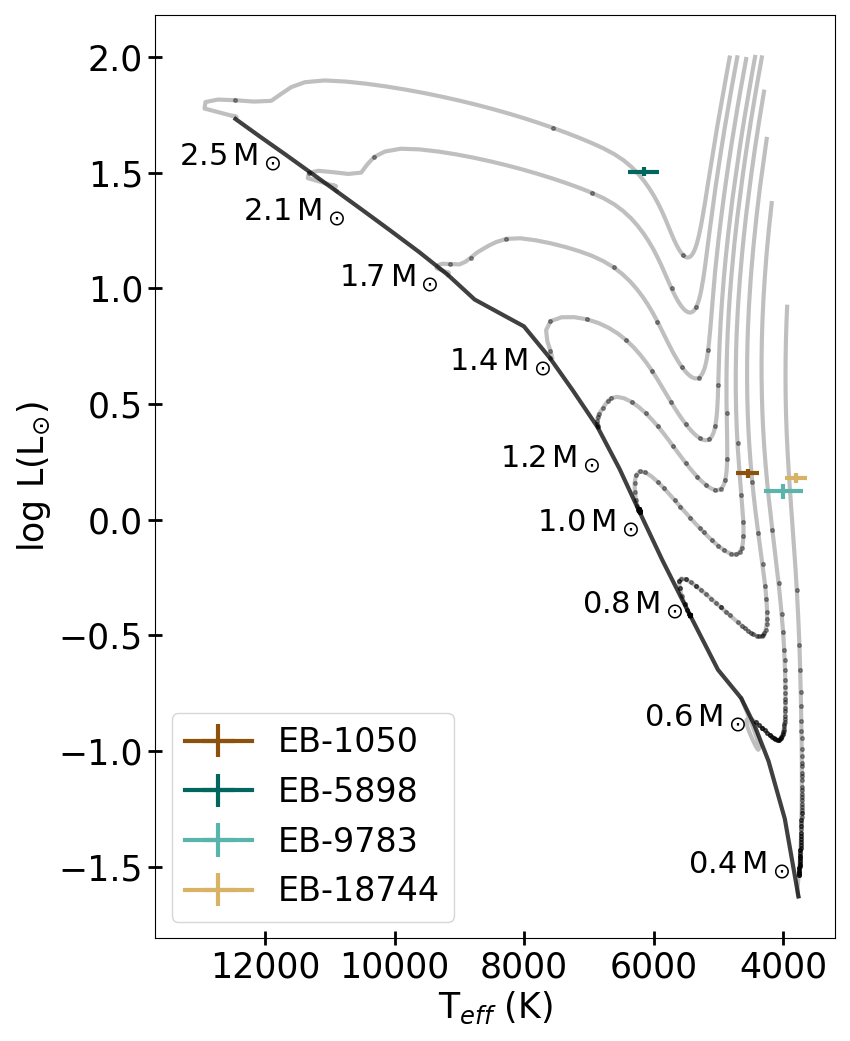}
    \caption{HRD of the subgiant secondaries in our four EBs. We overlay MIST pre-MS evolutionary tracks and the zero-age MS. The dots along each track indicate 1 Myr of evolution. Three of our subgiants are measurably discrepant with the corresponding pre-MS tracks given their measured dynamical masses, and therefore they are most likely post-Algols.}
    \label{fig:HRD2}
\end{figure}

We present an HRD of our subgiant secondaries in Figure~\ref{fig:HRD2}. We overlay MIST pre-MS evolutionary tracks \citep{Choi2016} to determine if our systems are post-Algols or nascent EBs. As found in \citet{Moe2015a}, the subgiant in EB-5898 is substantially more luminous and hotter than solar-mass pre-MS stars. According to the pre-MS tracks, the subgiant secondary in EB-5898 would be 2.5\,\Msun\ and 1.5 Myr old if it were a pre-MS star. However, the inferred pre-MS mass of 2.5\,\Msun\ is highly discrepant with its measured dynamical mass of 1.08\,$\pm$\,0.16\,\Msun. Moreover, the B-type primary is evolved with log\,$g$ = 3.75 and corresponding age of $\tau$ $\approx$ 35 Myr (see Fig.~\ref{fig:HRDPrimary}). The B-type primary cannot be coeval with a 1.5\,Myr old pre-MS companion. We therefore conclude that EB-5898 is not a nascent EB but instead a post-Algol with high probability. 

The subgiant secondaries in the other three systems lie along the Hayashi tracks of low-mass stars. The subgiants in EB-9783 and EB-18744 are slightly cooler than and also discrepant with their corresponding pre-MS tracks given their measured dynamical masses. Like EB-5898, the B-type primary in EB-9783 has log\,$g$ = 3.75, is more evolved (see Fig.~\ref{fig:HRDPrimary}), and cannot be coeval with a pre-MS companion on the Hayashi track. We conclude that EB-18744 and especially EB-9783 are most likely post-Algols.

Conversely, the subgiant secondary in EB-1050 is consistent with its corresponding pre-MS track given its measured dynamical mass. Its B-type primary is also consistent with the ZAMS (see Fig.~\ref{fig:HRDPrimary}) and therefore could be coeval with a pre-MS companion on the Hayashi track. EB-1050 could possibly be a post-Algol, but it may  potentially be a runaway nascent EB that is only 1 Myr old yet far removed from its birth H\,II region (see Section~\ref{sec:Obj}).

\begin{deluxetable*}{lrrrrrrrrrrrr}[bt!]
    \setlength{\tabcolsep}{3pt}
    \tablecaption{Finalized System Parameters \label{tab:finalTable}}
    \tablehead{
        \colhead{} & \multicolumn{2}{c}{System} & \colhead{} & \multicolumn{4}{c}{Primary} & \colhead{} & \multicolumn{4}{c}{Secondary} \\
        \colhead{Object} & \colhead{$P$(days)} & \colhead{$i$(\textdegree)} & \colhead{} & \colhead{$M_\mathrm{1}$(\Msun)} & \colhead{$T_\mathrm{eff, 1}$(K)} & \colhead{$R_\mathrm{1}$(\Rsun)} & \colhead{$RLFF_{\mathrm{1}}$} & \colhead{} & \colhead{$M_\mathrm{2}$(\Msun)} & \colhead{$T_\mathrm{eff, 2}$(K)} & \colhead{$R_\mathrm{2}$(\Rsun)} & \colhead{$RLFF_{\mathrm{2}}$} 
    }
    \startdata
    EB-1050 & 1.670784 & 84.8 & & 7.1 & 22,000 & 3.37 & 0.52 & & 1.08 & 4550 & 2.03 & 0.73 \\
     & $\pm$\,0.000003 & $\pm$\,0.9 & & $\pm$\,0.6 & $\pm$\,800 & $\pm$\,0.10 & $\pm$\,0.03 & & $\pm$\,0.09 & $\pm$\,180 & $\pm$\,0.07 & $\pm$\,0.03 \\
    \hline
    EB-5898 & 5.323882 & 83.1 & & 7.7 & 21,000 & 4.87 & 0.33 & & 1.08 & 6150 & 4.98 & 0.83 \\
     & $\pm$\,0.000007 & $\pm$\,0.6 & & $\pm$\,0.7 & $\pm$\,800 & $\pm$\,0.16 & $\pm$\,0.03 & & $\pm$\,0.16 & $\pm$\,240 & $\pm$\,0.17 & $\pm$\,0.03 \\
    \hline 
    EB-9783 & 1.751516 & 70.5 & & 6.3 & 18,700 & 3.43 & 0.52 & & 0.91 & 4000 & 2.40 & 0.89 \\
     & $\pm$\,0.000012 & $\pm$\,0.6 & & $\pm$\,0.6 & $\pm$\,800 & $\pm$\,0.10 & $\pm$\,0.03 & & $\pm$\,0.08 & $\pm$\,300 & $\pm$\,0.08 & $\pm$\,0.03 \\
    \hline
    EB-18744 & 2.073373 & 75.7 & & 6.9 & 21,200 & 3.28 & 0.45 & & 1.22 & 3800 & 2.84 & 0.85 \\
      & $\pm$\,0.000003 & $\pm$\,0.6 & & $\pm$\,0.6 & $\pm$\,800 & $\pm$\,0.10 & $\pm$\,0.03 & & $\pm$\,0.10 & $\pm$\,170 & $\pm$\,0.09 & $\pm$\,0.03 \\
    \enddata
\end{deluxetable*}

According to binary evolution models, a subgiant donor detaches from its Roche lobe between Case A and AB mass transfer and contracts to $RLFF_2$ = 75\% during its TAMS \citep[see second panel of Fig.~2 in][]{Sen2022}. After Case AB mass transfer, the donor star rapidly shrinks on a thermal timescale into a WD. The subgiant secondaries in our post-Algols span $RLFF_2$ = 73\%\,-\,89\%, discrepant with unity at the $>$4$\sigma$ level. Our EBs are unlikely to be observed during the extremely short-lived phase between Case AB mass transfer and a WD (see discussion in Section~\ref{sec:post-Algol}). We instead conclude that they are detached post-Algols evolving on a nuclear timescale observed between Case A and AB mass transfer. The measured subgiant Roche-lobe fill factors in our post-Algol candidates are consistent with binary evolution models. The discovery of our post-Algol candidates provides the first empirical evidence that intermediate-mass stars contract along the TAMS, confirming a sixty year prediction \citep{Iben1967a,Iben1967b}. 

For convenience, we summarize the finalized system parameters in Table~\ref{tab:finalTable}. We compute a weighted average for the orbital periods based on the OGLE-III and OGLE-IV analytic light curve fits. We adopt the radii from the PHOEBE light curve fits, which result in smaller uncertainties. 

The properties of EB-5898 are different compared to the other three post-Algol candidates. The subgiant secondary in EB-5898 is warmer and more luminous, the orbital period is greater, and the radii of both components are larger. EB-5898 likely had different initial period and masses than the other three systems in our study. Specifically, we speculate that the progenitor of EB-5898 had a slightly longer initial orbital period and comparable MS component masses. Mass transfer tends to be more conservative when the initial mass of the accretor is similar to the initial mass of the donor \citep{deMink2007,Schneider2015,Lechien2025}. Conservative Case A mass transfer more readily widens the orbit compared to non-conservative Case A mass transfer (see bottom panel of Fig.~2 in \citealt{Sen2022}). Since the orbital period is longer in EB-5898, the former donor is larger, warmer, and more luminous. Meanwhile, the progenitors of EB-1050, 9783, and 18744 possibly had more extreme initial mass ratios. Such low-mass accretors lead to highly non-conservative mass transfer, which only marginally widens the orbit. The observed periods $P$ = 1.7\,-\,2.1~days of post-Algols EB-1050, 9783, and 18744 are likely very similar to their initial periods if they evolved through highly non-conservative mass transfer.

Future binary evolution studies should carefully model the properties of post-Algols, using our observed systems to anchor their simulations. Mass transfer efficiency during Case A mass transfer is a crucial free parameter in binary population synthesis studies \citep{deMink2007,Deschamps2013,Siess2018,Sen2022,Lechien2025}. Whether mass transfer is conservative or significantly non-conservative leads to dramatic differences in the predicted final periods and masses of evolved binaries. Our post-Algol candidates can be used to constrain the mass transfer efficiency in binary evolution models. 

Regarding the subsequent evolution of our systems, the subgiants will re-expand and refill their Roche lobes, transferring mass to the mid-B primaries. The subgiants will quickly lose their hydrogen envelopes and their photospheres will contract into WDs. For three of our systems (EB-1050, 9783, and 18744), the mid-B primaries will gradually expand on the MS and fill their own Roche lobes, likely leading to a common envelope (CE) evolution due to the extreme mass ratios $q$ = $M_{\rm WD}$/$M_{\rm B}$ = 0.1\,-\,0.2. Depending on the efficiency of CE ejection, the systems will either merge or produce tight post-CE binary WDs. 

Meanwhile, the mid-B primary in EB-5898 is currently $M_1$ = 7.7\,$\pm$\,0.7 \Msun. It will gain additional mass during Case AB mass transfer, likely pushing it above $M_1$ $>$ 8\,\Msun,  possibly sufficient to explode as a supernova. Single stars with $M$ $>$ 8\,\Msun\ explode as supernovae within 50~Myr after birth, but rejuvenated mass gainers in evolved binaries can produce delayed supernovae after 50~Myr \citep{Zapartas2017}. The mid-B primary in our post-Algol EB-5898 could potentially be the progenitor of such a delayed supernova. Moreover, such a 8\,\Msun\ progenitor may possibly produce an electron capture supernova similar to SN\,1054 that created the Crab Nebula \citep{Nomoto1982,Nomoto1984,Smith2013}. 

\section{Summary}
\label{sec:summary}

By searching the OGLE-III catalog of EBs in the LMC, we identified 5 post-Algol candidates caught between Case A and Case AB mass transfer. Their light curves feature large reflection effect amplitudes $\Delta I_{\rm Refl}$ $>$ 0.08 mag and no ellipsoidal variability, indicating a detached configuration. The post-Algol candidates reside in old environments far removed from any H\,II regions. Given our sample of 86 semi-detached Algols, we expect to find a few detached post-Algols according to binary evolution models.  

We obtained multi-epoch echelle spectra of four of our post-Algol candidates with the MIKE spectrograph on the 6.5m Magellan-Clay telescope. We fit TLUSTY atmospheric models to the spectra. The primaries are mid-B dwarfs with surface gravities log\,$g$ = 3.75\,-\,4.00, effective temperatures $T_{\rm eff,1}$ = 18,700\,-\,22,000\,K, modest projected rotational velocities $v$\,sin\,$i$ = 86\,-\,131\,km\,s$^{-1}$, and LMC abundances ($\nicefrac{1}{2}$\,Z$_{\odot}$). By fitting the SEDs of the B-dwarf primaries, we measured their bolometric luminosities $L_1$. According to their locations on the HRD, the primary masses span $M_1$ = 6.3\,-\,7.7\,\Msun. From the multi-epoch radial velocities, we measured the dynamical masses of the subgiant secondaries to be $M_2$ = 0.91\,-\,1.22\,\Msun. By fitting PHOEBE models to the OGLE-III and OGLE-IV light curves, we measured the radii $R_1$ and $R_2$ of both components, the secondary effective temperatures $T_{\rm eff,2}$, the secondary albedos $A_2$, and the orbital inclinations $i$. The measured albedos span $A_2$ = 45\%\,-\,60\%, as expected for cool subgiants with convective envelopes. The measured Roche-lobe fill factors span $RLFF_2$ = 73\%\,-\,89\%, consistent with binary evolution models of post-Algols.

The prototype EB-5898 is at a longer orbital period ($P$ = 5.3~days) and has a more luminous and hotter subgiant secondary compared to the other three systems. EB-5898 cannot be a nascent EB. If it had a pre-MS companion, the mass inferred from pre-MS tracks would be highly discrepant with its measured dynamical mass and the inferred age would not be coeval with its evolved B-type primary. EB-5898 is most likely a post-Algol with a subgiant on the TAMS caught between Case A and Case AB mass transfer. The other three candidates have shorter orbital periods ($P$ = 1.7\,-\,2.1~days) and cooler, less luminous secondaries. For EB-9783 and EB-18744, the companion masses inferred from pre-MS tracks are slightly discrepant with their measured dynamical masses. The primary B-dwarf in EB-9783 is more evolved and cannot be coeval with a pre-MS companion. Thus EB-18744 and especially EB-9783 are likely not nascent EBs but most probably post-Algols. The final system EB-1050 has a primary B-dwarf near the ZAMS and a measured companion mass that is consistent with its corresponding pre-MS track. EB-1050 may be a post-Algol but could also potentially be a 1 Myr old runaway nascent EB that has been kicked tens of parsecs from its birth H\,II region. 

Donor stars can detach from their Roche lobes while not on the TAMS, but such evolution occurs on a rapid thermal timescale. Contraction along the TAMS occurs on a longer nuclear timescale, lasting $\approx$5\% of the mass-transferring phases. At least some of our post-Algol candidates are probably genuine and contain subgiant secondaries that have contracted on the TAMS. The discovery of detached post-Algols provides the first empirical evidence that intermediate-mass stars contract on the TAMS, confirming a 60-year prediction.

\facilities{This work made use of the Magellan Inamori Kyocera Echelle (MIKE) spectrograph attached to the Magellan 2 - Clay Telescope at the Las Campanas Observatory.}

\begin{acknowledgements}

We acknowledge financial support from NASA Grant \#80NSSC25M7130. Computations were performed using the University of Wyoming (UW) Advance Research Computing Center MedicineBow HPC, a UW managed computational resource available to UW researchers including faculty, staff, students, and collaborators.

\end{acknowledgements}

\bibliographystyle{aasjournalv7}  
\bibliography{bib}

@INPROCEEDINGS{Bernstein2003,
       author = {{Bernstein}, Rebecca and {Shectman}, Stephen A. and {Gunnels}, Steven M. and {Mochnacki}, Stefan and {Athey}, Alex E.},
        title = "{MIKE: A Double Echelle Spectrograph for the Magellan Telescopes at Las Campanas Observatory}",
    booktitle = {Instrument Design and Performance for Optical/Infrared Ground-based Telescopes},
         year = 2003,
       editor = {{Iye}, Masanori and {Moorwood}, Alan F.~M.},
       series = {Society of Photo-Optical Instrumentation Engineers (SPIE) Conference Series},
       volume = {4841},
        month = mar,
        pages = {1694-1704},
          doi = {10.1117/12.461502},
       adsurl = {https://ui.adsabs.harvard.edu/abs/2003SPIE.4841.1694B}
}

@ARTICLE{Baron2012,
       author = {{Baron}, F. and {Monnier}, J.~D. and {Pedretti}, E. and {Zhao}, M. and {Schaefer}, G. and {Parks}, R. and {Che}, X. and {Thureau}, N. and {ten Brummelaar}, T.~A. and {McAlister}, H.~A. and {Ridgway}, S.~T. and {Farrington}, C. and {Sturmann}, J. and {Sturmann}, L. and {Turner}, N.},
        title = "{Imaging the Algol Triple System in the H Band with the CHARA Interferometer}",
      journal = {\apj},
         year = 2012,
        month = jun,
       volume = {752},
       number = {1},
          eid = {20},
        pages = {20},
          doi = {10.1088/0004-637X/752/1/20},
archivePrefix = {arXiv},
       eprint = {1205.0754},
 primaryClass = {astro-ph.SR},
       adsurl = {https://ui.adsabs.harvard.edu/abs/2012ApJ...752...20B}
}

@ARTICLE{Richter1987,
       author = {{Richter}, O. -G. and {Tammann}, G.~A. and {Huchtmeier}, W.~K.},
        title = "{HI observations of galaxies in a catalog of nearby galaxies. II. The motion of the sun and the galaxy and the velocity dispersion of 'field' galaxies.}",
      journal = {\aap},
         year = 1987,
        month = jan,
       volume = {171},
        pages = {33-40},
       adsurl = {https://ui.adsabs.harvard.edu/abs/1987A&A...171...33R}
}

@ARTICLE{Hoogerwerf2001,
       author = {{Hoogerwerf}, R. and {de Bruijne}, J.~H.~J. and {de Zeeuw}, P.~T.},
        title = "{On the origin of the O and B-type stars with high velocities. II. Runaway stars and pulsars ejected from the nearby young stellar groups}",
      journal = {\aap},
         year = 2001,
        month = jan,
       volume = {365},
        pages = {49-77},
          doi = {10.1051/0004-6361:20000014},
archivePrefix = {arXiv},
       eprint = {astro-ph/0010057},
 primaryClass = {astro-ph},
       adsurl = {https://ui.adsabs.harvard.edu/abs/2001A&A...365...49H}
}

@ARTICLE{Gies1986,
       author = {{Gies}, D.~R. and {Bolton}, C.~T.},
        title = "{The Binary Frequency and Origin of the OB Runaway Stars}",
      journal = {\apjs},
         year = 1986,
        month = jun,
       volume = {61},
        pages = {419},
          doi = {10.1086/191118},
       adsurl = {https://ui.adsabs.harvard.edu/abs/1986ApJS...61..419G}
}

@ARTICLE{Fujii2011,
       author = {{Fujii}, Michiko S. and {Portegies Zwart}, Simon},
        title = "{The Origin of OB Runaway Stars}",
      journal = {Science},
         year = 2011,
        month = dec,
       volume = {334},
       number = {6061},
        pages = {1380},
          doi = {10.1126/science.1211927},
archivePrefix = {arXiv},
       eprint = {1111.3644},
 primaryClass = {astro-ph.GA},
       adsurl = {https://ui.adsabs.harvard.edu/abs/2011Sci...334.1380F}
}

@ARTICLE{deBurgos2025,
       author = {{de Burgos}, A. and {Sim{\'o}n-D{\'\i}az}, S. and {Urbaneja}, M.~A. and {Holgado}, G. and {Ekstr{\"o}m}, S. and {Ram{\'\i}rez-Tannus}, M.~C. and {Zari}, E.},
        title = "{The IACOB project: XIV. New clues on the location of the TAMS in the massive star domain}",
      journal = {\aap},
         year = 2025,
        month = mar,
       volume = {695},
          eid = {A87},
        pages = {A87},
          doi = {10.1051/0004-6361/202453242},
archivePrefix = {arXiv},
       eprint = {2501.17984},
 primaryClass = {astro-ph.SR},
       adsurl = {https://ui.adsabs.harvard.edu/abs/2025A&A...695A..87D}
}

@ARTICLE{Dotter2008,
       author = {{Dotter}, Aaron and {Chaboyer}, Brian and {Jevremovi{\'c}}, Darko and {Kostov}, Veselin and {Baron}, E. and {Ferguson}, Jason W.},
        title = "{The Dartmouth Stellar Evolution Database}",
      journal = {\apjs},
         year = 2008,
        month = sep,
       volume = {178},
       number = {1},
        pages = {89-101},
          doi = {10.1086/589654},
archivePrefix = {arXiv},
       eprint = {0804.4473},
 primaryClass = {astro-ph},
       adsurl = {https://ui.adsabs.harvard.edu/abs/2008ApJS..178...89D}
}

@ARTICLE{Bressan2012,
       author = {{Bressan}, Alessandro and {Marigo}, Paola and {Girardi}, L{\'e}o. and {Salasnich}, Bernardo and {Dal Cero}, Claudia and {Rubele}, Stefano and {Nanni}, Ambra},
        title = "{PARSEC: stellar tracks and isochrones with the PAdova and TRieste Stellar Evolution Code}",
      journal = {\mnras},
         year = 2012,
        month = nov,
       volume = {427},
       number = {1},
        pages = {127-145},
          doi = {10.1111/j.1365-2966.2012.21948.x},
archivePrefix = {arXiv},
       eprint = {1208.4498},
 primaryClass = {astro-ph.SR},
       adsurl = {https://ui.adsabs.harvard.edu/abs/2012MNRAS.427..127B}
}

@ARTICLE{Choi2016,
       author = {{Choi}, Jieun and {Dotter}, Aaron and {Conroy}, Charlie and {Cantiello}, Matteo and {Paxton}, Bill and {Johnson}, Benjamin D.},
        title = "{Mesa Isochrones and Stellar Tracks (MIST). I. Solar-scaled Models}",
      journal = {\apj},
         year = 2016,
        month = jun,
       volume = {823},
       number = {2},
          eid = {102},
        pages = {102},
          doi = {10.3847/0004-637X/823/2/102},
archivePrefix = {arXiv},
       eprint = {1604.08592},
 primaryClass = {astro-ph.SR},
       adsurl = {https://ui.adsabs.harvard.edu/abs/2016ApJ...823..102C}
}

@ARTICLE{Iben1967a,
       author = {{Iben}, Jr., Icko},
        title = "{Stellar Evolution Within and off the Main Sequence}",
      journal = {\araa},
         year = 1967,
        month = jan,
       volume = {5},
        pages = {571},
          doi = {10.1146/annurev.aa.05.090167.003035},
       adsurl = {https://ui.adsabs.harvard.edu/abs/1967ARA&A...5..571I}
}

@ARTICLE{Iben1967b,
       author = {{Iben}, Jr., Icko},
        title = "{Stellar Evolution. VII. The Evolution of a 2.25 M\_\{sun\} Star from the Main Sequence to the Helium-Burning Phase}",
      journal = {\apj},
         year = 1967,
        month = feb,
       volume = {147},
        pages = {650},
          doi = {10.1086/149041},
       adsurl = {https://ui.adsabs.harvard.edu/abs/1967ApJ...147..650I}
}

@ARTICLE{Pigulski2024,
       author = {{Pigulski}, A.},
        title = "{BRITE nascent binaries}",
      journal = {Bulletin de la Societe Royale des Sciences de Liege},
         year = 2024,
        month = dec,
       volume = {93},
       number = {3},
        pages = {4-20},
          doi = {10.25518/0037-9565.12243},
archivePrefix = {arXiv},
       eprint = {2410.19120},
 primaryClass = {astro-ph.SR},
       adsurl = {https://ui.adsabs.harvard.edu/abs/2024BSRSL..93....4P}
}

@ARTICLE{Jerzykiewicz2021,
       author = {{Jerzykiewicz}, M. and {Pigulski}, A. and {Michalska}, G. and {Mo{\'z}dzierski}, D. and {Ratajczak}, M. and {Handler}, G. and {Moffat}, A.~F.~J. and {Pablo}, H. and {Popowicz}, A. and {Wade}, G.~A. and {Zwintz}, K.},
        title = "{BRITE observations of {\ensuremath{\nu}} Centauri and {\ensuremath{\gamma}} Lupi, the first non-eclipsing members of the new class of nascent binaries}",
      journal = {\mnras},
         year = 2021,
        month = jun,
       volume = {503},
       number = {4},
        pages = {5554-5568},
          doi = {10.1093/mnras/stab846},
archivePrefix = {arXiv},
       eprint = {2104.11770},
 primaryClass = {astro-ph.SR},
       adsurl = {https://ui.adsabs.harvard.edu/abs/2021MNRAS.503.5554J}
}

@ARTICLE{BrownSevilla2021,
       author = {{Brown-Sevilla}, S.~B. and {Nascimbeni}, V. and {Borsato}, L. and {Tartaglia}, L. and {Nardiello}, D. and {Granata}, V. and {Libralato}, M. and {Damasso}, M. and {Piotto}, G. and {Pollacco}, D. and {West}, R.~G. and {Colombo}, L.~S. and {Cunial}, A. and {Piazza}, G. and {Scaggiante}, F.},
        title = "{A new photometric and dynamical study of the eclipsing binary star HW Virginis}",
      journal = {\mnras},
         year = 2021,
        month = sep,
       volume = {506},
       number = {2},
        pages = {2122-2135},
          doi = {10.1093/mnras/stab1843},
archivePrefix = {arXiv},
       eprint = {2106.15632},
 primaryClass = {astro-ph.SR},
       adsurl = {https://ui.adsabs.harvard.edu/abs/2021MNRAS.506.2122B}
}

@ARTICLE{Prsa2011,
       author = {{Pr{\v{s}}a}, Andrej and {Pepper}, Joshua and {Stassun}, Keivan G.},
        title = "{Expected Large Synoptic Survey Telescope (LSST) Yield of Eclipsing Binary Stars}",
      journal = {\aj},
         year = 2011,
        month = aug,
       volume = {142},
       number = {2},
          eid = {52},
        pages = {52},
          doi = {10.1088/0004-6256/142/2/52},
archivePrefix = {arXiv},
       eprint = {1105.6011},
 primaryClass = {astro-ph.SR},
       adsurl = {https://ui.adsabs.harvard.edu/abs/2011AJ....142...52P}
}

@ARTICLE{Nomoto1982,
       author = {{Nomoto}, Ken'ichi and {Sparks}, Warren M. and {Fesen}, Robert A. and {Gull}, Theodore R. and {Miyaji}, S. and {Sugimoto}, D.},
        title = "{The Crab Nebula's progenitor}",
      journal = {\nat},
         year = 1982,
        month = oct,
       volume = {299},
       number = {5886},
        pages = {803-805},
          doi = {10.1038/299803a0},
       adsurl = {https://ui.adsabs.harvard.edu/abs/1982Natur.299..803N}
}

@ARTICLE{Nomoto1984,
       author = {{Nomoto}, K.},
        title = "{Evolution of 8-10 solar mass stars toward electron capture supernovae. I - Formation of electron-degenerate O + NE + MG cores.}",
      journal = {\apj},
         year = 1984,
        month = feb,
       volume = {277},
        pages = {791-805},
          doi = {10.1086/161749},
       adsurl = {https://ui.adsabs.harvard.edu/abs/1984ApJ...277..791N}
}

@ARTICLE{Smith2013,
       author = {{Smith}, Nathan},
        title = "{The Crab nebula and the class of Type IIn-P supernovae caused by sub-energetic electron-capture explosions}",
      journal = {\mnras},
         year = 2013,
        month = sep,
       volume = {434},
       number = {1},
        pages = {102-113},
          doi = {10.1093/mnras/stt1004},
archivePrefix = {arXiv},
       eprint = {1304.0689},
 primaryClass = {astro-ph.HE},
       adsurl = {https://ui.adsabs.harvard.edu/abs/2013MNRAS.434..102S}
}

@ARTICLE{Schneider2015,
       author = {{Schneider}, F.~R.~N. and {Izzard}, R.~G. and {Langer}, N. and {de Mink}, S.~E.},
        title = "{Evolution of Mass Functions of Coeval Stars through Wind Mass Loss and Binary Interactions}",
      journal = {\apj},
         year = 2015,
        month = may,
       volume = {805},
       number = {1},
          eid = {20},
        pages = {20},
          doi = {10.1088/0004-637X/805/1/20},
archivePrefix = {arXiv},
       eprint = {1504.01735},
 primaryClass = {astro-ph.SR},
       adsurl = {https://ui.adsabs.harvard.edu/abs/2015ApJ...805...20S}
}

@ARTICLE{deMink2007,
       author = {{de Mink}, S.~E. and {Pols}, O.~R. and {Hilditch}, R.~W.},
        title = "{Efficiency of mass transfer in massive close binaries. Tests from double-lined eclipsing binaries in the SMC}",
      journal = {\aap},
         year = 2007,
        month = jun,
       volume = {467},
       number = {3},
        pages = {1181-1196},
          doi = {10.1051/0004-6361:20067007},
archivePrefix = {arXiv},
       eprint = {astro-ph/0703480},
 primaryClass = {astro-ph},
       adsurl = {https://ui.adsabs.harvard.edu/abs/2007A&A...467.1181D}
}

@ARTICLE{Lechien2025,
       author = {{Lechien}, Thibault and {de Mink}, Selma E. and {Valli}, Ruggero and {Rubio}, Amanda C. and {van Son}, Lieke A.~C. and {Klement}, Robert and {Jin}, Harim and {Pols}, Onno},
        title = "{Binary Stars Take What They Get: Evidence for Efficient Mass Transfer from Stripped Stars with Rapidly Rotating Companions}",
      journal = {\apjl},
         year = 2025,
        month = sep,
       volume = {990},
       number = {2},
          eid = {L51},
        pages = {L51},
          doi = {10.3847/2041-8213/adfdd4},
archivePrefix = {arXiv},
       eprint = {2505.14780},
 primaryClass = {astro-ph.SR},
       adsurl = {https://ui.adsabs.harvard.edu/abs/2025ApJ...990L..51L}
}

@ARTICLE{Petrovic2005,
       author = {{Petrovic}, J. and {Langer}, N. and {van der Hucht}, K.~A.},
        title = "{Constraining the mass transfer in massive binaries through progenitor evolution models of Wolf-Rayet+O binaries}",
      journal = {\aap},
         year = 2005,
        month = jun,
       volume = {435},
       number = {3},
        pages = {1013-1030},
          doi = {10.1051/0004-6361:20042368},
archivePrefix = {arXiv},
       eprint = {astro-ph/0504242},
 primaryClass = {astro-ph},
       adsurl = {https://ui.adsabs.harvard.edu/abs/2005A&A...435.1013P}
}

@ARTICLE{Wellstein2001,
       author = {{Wellstein}, S. and {Langer}, N. and {Braun}, H.},
        title = "{Formation of contact in massive close binaries}",
      journal = {\aap},
         year = 2001,
        month = apr,
       volume = {369},
        pages = {939-959},
          doi = {10.1051/0004-6361:20010151},
archivePrefix = {arXiv},
       eprint = {astro-ph/0102244},
 primaryClass = {astro-ph},
       adsurl = {https://ui.adsabs.harvard.edu/abs/2001A&A...369..939W}
}

@ARTICLE{Schurmann2024,
       author = {{Sch{\"u}rmann}, C. and {Langer}, N. and {Kramer}, J.~A. and {Marchant}, P. and {Wang}, C. and {Sen}, K.},
        title = "{Analytic approximations for massive close post-mass transfer binary systems}",
      journal = {\aap},
         year = 2024,
        month = oct,
       volume = {690},
          eid = {A282},
        pages = {A282},
          doi = {10.1051/0004-6361/202450353},
archivePrefix = {arXiv},
       eprint = {2404.08612},
 primaryClass = {astro-ph.SR},
       adsurl = {https://ui.adsabs.harvard.edu/abs/2024A&A...690A.282S}
}

@ARTICLE{Mowlavi2023,
       author = {{Mowlavi}, N. and {Holl}, B. and {Lecoeur-Ta{\"\i}bi}, I. and {Barblan}, F. and {Kochoska}, A. and {Pr{\v{s}}a}, A. and {Mazeh}, T. and {Rimoldini}, L. and {Gavras}, P. and {Audard}, M. and {Jevardat de Fombelle}, G. and {Nienartowicz}, K. and {Garc{\'\i}a-Lario}, P. and {Eyer}, L.},
        title = "{Gaia Data Release 3. The first Gaia catalogue of eclipsing-binary candidates}",
      journal = {\aap},
         year = 2023,
        month = jun,
       volume = {674},
          eid = {A16},
        pages = {A16},
          doi = {10.1051/0004-6361/202245330},
archivePrefix = {arXiv},
       eprint = {2211.00929},
 primaryClass = {astro-ph.SR},
       adsurl = {https://ui.adsabs.harvard.edu/abs/2023A&A...674A..16M}
}

@ARTICLE{Glowacki2024,
       author = {{G{\l}owacki}, M. and {Soszy{\'n}ski}, I. and {Udalski}, A. and {Szyma{\'n}ski}, M.~K. and {Skowron}, J. and {Skowron}, D.~M. and {Mr{\'o}z}, P. and {Pietrukowicz}, P. and {Poleski}, R. and {Koz{\l}owski}, S. and {Iwanek}, P. and {Wrona}, M. and {Ulaczyk}, K. and {Rybicki}, K. and {Gromadzki}, M. and {Mr{\'o}z}, M.~J. and {Urbanowicz}, M.},
        title = "{The OGLE Collection of Variable Stars. Over 75 000 Eclipsing and Ellipsoidal Binary Systems in the Magellanic Clouds}",
      journal = {\actaa},
         year = 2024,
        month = dec,
       volume = {74},
       number = {4},
        pages = {241-264},
          doi = {10.32023/0001-5237/74.4.1},
archivePrefix = {arXiv},
       eprint = {2503.15596},
 primaryClass = {astro-ph.SR},
       adsurl = {https://ui.adsabs.harvard.edu/abs/2024AcA....74..241G}
}

@ARTICLE{Soszynski2016,
       author = {{Soszy{\'n}ski}, I. and {Pawlak}, M. and {Pietrukowicz}, P. and {Udalski}, A. and {Szyma{\'n}ski}, M.~K. and {Wyrzykowski}, {\L}. and {Ulaczyk}, K. and {Poleski}, R. and {Koz{\l}owski}, S. and {Skowron}, D.~M. and {Skowron}, J. and {Mr{\'o}z}, P. and {Hamanowicz}, A.},
        title = "{The OGLE Collection of Variable Stars. Over 450 000 Eclipsing and Ellipsoidal Binary Systems Toward the Galactic Bulge}",
      journal = {\actaa},
         year = 2016,
        month = dec,
       volume = {66},
       number = {4},
        pages = {405-420},
          doi = {10.48550/arXiv.1701.03105},
archivePrefix = {arXiv},
       eprint = {1701.03105},
 primaryClass = {astro-ph.SR},
       adsurl = {https://ui.adsabs.harvard.edu/abs/2016AcA....66..405S}
}

@INPROCEEDINGS{Offner2023,
       author = {{Offner}, S.~S.~R. and {Moe}, M. and {Kratter}, K.~M. and {Sadavoy}, S.~I. and {Jensen}, E.~L.~N. and {Tobin}, J.~J.},
        title = "{The Origin and Evolution of Multiple Star Systems}",
    booktitle = {Protostars and Planets VII},
         year = 2023,
       editor = {{Inutsuka}, S. and {Aikawa}, Y. and {Muto}, T. and {Tomida}, K. and {Tamura}, M.},
       series = {Astronomical Society of the Pacific Conference Series},
       volume = {534},
        month = jul,
        pages = {275},
          doi = {10.48550/arXiv.2203.10066},
archivePrefix = {arXiv},
       eprint = {2203.10066},
 primaryClass = {astro-ph.SR},
       adsurl = {https://ui.adsabs.harvard.edu/abs/2023ASPC..534..275O}
}

@ARTICLE{Lanz2007,
       author = {{Lanz}, Thierry and {Hubeny}, Ivan},
        title = "{A Grid of NLTE Line-blanketed Model Atmospheres of Early B-Type Stars}",
      journal = {\apjs},
         year = 2007,
        month = mar,
       volume = {169},
       number = {1},
        pages = {83-104},
          doi = {10.1086/511270},
archivePrefix = {arXiv},
       eprint = {astro-ph/0611891},
 primaryClass = {astro-ph},
       adsurl = {https://ui.adsabs.harvard.edu/abs/2007ApJS..169...83L}
}

@ARTICLE{Pietrzynski2019,
       author = {{Pietrzy{\'n}ski}, G. and {Graczyk}, D. and {Gallenne}, A. and {Gieren}, W. and {Thompson}, I.~B. and {Pilecki}, B. and {Karczmarek}, P. and {G{\'o}rski}, M. and {Suchomska}, K. and {Taormina}, M. and {Zgirski}, B. and {Wielg{\'o}rski}, P. and {Ko{\l}aczkowski}, Z. and {Konorski}, P. and {Villanova}, S. and {Nardetto}, N. and {Kervella}, P. and {Bresolin}, F. and {Kudritzki}, R.~P. and {Storm}, J. and {Smolec}, R. and {Narloch}, W.},
        title = "{A distance to the Large Magellanic Cloud that is precise to one per cent}",
      journal = {\nat},
         year = 2019,
        month = mar,
       volume = {567},
       number = {7747},
        pages = {200-203},
          doi = {10.1038/s41586-019-0999-4},
archivePrefix = {arXiv},
       eprint = {1903.08096},
 primaryClass = {astro-ph.GA},
       adsurl = {https://ui.adsabs.harvard.edu/abs/2019Natur.567..200P}
}

@ARTICLE{Kelson2000,
       author = {{Kelson}, Daniel D. and {Illingworth}, Garth D. and {van Dokkum}, Pieter G. and {Franx}, Marijn},
        title = "{The Evolution of Early-Type Galaxies in Distant Clusters. II. Internal Kinematics of 55 Galaxies in the z=0.33 Cluster CL 1358+62}",
      journal = {\apj},
         year = 2000,
        month = mar,
       volume = {531},
       number = {1},
        pages = {159-183},
          doi = {10.1086/308445},
archivePrefix = {arXiv},
       eprint = {astro-ph/9908257},
 primaryClass = {astro-ph},
       adsurl = {https://ui.adsabs.harvard.edu/abs/2000ApJ...531..159K}
}

@ARTICLE{Kelson2003,
       author = {{Kelson}, Daniel D.},
        title = "{Optimal Techniques in Two-dimensional Spectroscopy: Background Subtraction for the 21st Century}",
      journal = {\pasp},
         year = 2003,
        month = jun,
       volume = {115},
       number = {808},
        pages = {688-699},
          doi = {10.1086/375502},
archivePrefix = {arXiv},
       eprint = {astro-ph/0303507},
 primaryClass = {astro-ph},
       adsurl = {https://ui.adsabs.harvard.edu/abs/2003PASP..115..688K}
}

@ARTICLE{Prsa2005,
       author = {{Pr{\v{s}}a}, A. and {Zwitter}, T.},
        title = "{A Computational Guide to Physics of Eclipsing Binaries. I. Demonstrations and Perspectives}",
      journal = {\apj},
         year = 2005,
        month = jul,
       volume = {628},
       number = {1},
        pages = {426-438},
          doi = {10.1086/430591},
archivePrefix = {arXiv},
       eprint = {astro-ph/0503361},
 primaryClass = {astro-ph},
       adsurl = {https://ui.adsabs.harvard.edu/abs/2005ApJ...628..426P}
}

@ARTICLE{Fitzpatrick1999,
       author = {{Fitzpatrick}, Edward L.},
        title = "{Correcting for the Effects of Interstellar Extinction}",
      journal = {\pasp},
         year = 1999,
        month = jan,
       volume = {111},
       number = {755},
        pages = {63-75},
          doi = {10.1086/316293},
archivePrefix = {arXiv},
       eprint = {astro-ph/9809387},
 primaryClass = {astro-ph},
       adsurl = {https://ui.adsabs.harvard.edu/abs/1999PASP..111...63F}
}

@ARTICLE{Zaritsky2004,
       author = {{Zaritsky}, Dennis and {Harris}, Jason and {Thompson}, Ian B. and {Grebel}, Eva K.},
        title = "{The Magellanic Clouds Photometric Survey: The Large Magellanic Cloud Stellar Catalog and Extinction Map}",
      journal = {\aj},
         year = 2004,
        month = oct,
       volume = {128},
       number = {4},
        pages = {1606-1614},
          doi = {10.1086/423910},
archivePrefix = {arXiv},
       eprint = {astro-ph/0407006},
 primaryClass = {astro-ph},
       adsurl = {https://ui.adsabs.harvard.edu/abs/2004AJ....128.1606Z}
}

@ARTICLE{Chen2022,
       author = {{Chen}, B. -Q. and {Guo}, H. -L. and {Gao}, J. and {Yang}, M. and {Liu}, Y. -L. and {Jiang}, B. -W.},
        title = "{Dust distributions in the magellanic clouds}",
      journal = {\mnras},
         year = 2022,
        month = mar,
       volume = {511},
       number = {1},
        pages = {1317-1329},
          doi = {10.1093/mnras/stac072},
archivePrefix = {arXiv},
       eprint = {2201.03152},
 primaryClass = {astro-ph.GA},
       adsurl = {https://ui.adsabs.harvard.edu/abs/2022MNRAS.511.1317C}
}

@ARTICLE{Udalski2008,
       author = {{Udalski}, A. and {Soszynski}, I. and {Szymanski}, M.~K. and {Kubiak}, M. and {Pietrzynski}, G. and {Wyrzykowski}, L. and {Szewczyk}, O. and {Ulaczyk}, K. and {Poleski}, R.},
        title = "{The Optical Gravitational Lensing Experiment. OGLE-III Photometric Maps of the Large Magellanic Cloud}",
      journal = {\actaa},
         year = 2008,
        month = jun,
       volume = {58},
        pages = {89-102},
          doi = {10.48550/arXiv.0807.3889},
archivePrefix = {arXiv},
       eprint = {0807.3889},
 primaryClass = {astro-ph},
       adsurl = {https://ui.adsabs.harvard.edu/abs/2008AcA....58...89U}
}

@ARTICLE{Gracyk2011,
       author = {{Graczyk}, D. and {Soszy{\'n}ski}, I. and {Poleski}, R. and {Pietrzy{\'n}ski}, G. and {Udalski}, A. and {Szyma{\'n}ski}, M.~K. and {Kubiak}, M. and {Wyrzykowski}, {\L}. and {Ulaczyk}, K.},
        title = "{The Optical Gravitational Lensing Experiment. The OGLE-III Catalog of Variable Stars. XII. Eclipsing Binary Stars in the Large Magellanic Cloud}",
      journal = {\actaa},
         year = 2011,
        month = jun,
       volume = {61},
       number = {2},
        pages = {103-122},
          doi = {10.48550/arXiv.1108.0446},
archivePrefix = {arXiv},
       eprint = {1108.0446},
 primaryClass = {astro-ph.SR},
       adsurl = {https://ui.adsabs.harvard.edu/abs/2011AcA....61..103G}
}

@ARTICLE{Moe2015b,
       author = {{Moe}, Maxwell and {Di~Stefano}, Rosanne},
        title = "{Early-type Eclipsing Binaries with Intermediate Orbital Periods}",
      journal = {\apj},
         year = 2015,
        month = sep,
       volume = {810},
       number = {1},
          eid = {61},
        pages = {61},
          doi = {10.1088/0004-637X/810/1/61},
archivePrefix = {arXiv},
       eprint = {1501.03152},
 primaryClass = {astro-ph.SR},
       adsurl = {https://ui.adsabs.harvard.edu/abs/2015ApJ...810...61M}
}

@ARTICLE{Moe2015a,
       author = {{Moe}, Maxwell and {Di~Stefano}, Rosanne},
        title = "{A New Class of Nascent Eclipsing Binaries with Extreme Mass Ratios}",
      journal = {\apj},
         year = 2015,
        month = mar,
       volume = {801},
       number = {2},
          eid = {113},
        pages = {113},
          doi = {10.1088/0004-637X/801/2/113},
archivePrefix = {arXiv},
       eprint = {1410.8138},
 primaryClass = {astro-ph.SR},
       adsurl = {https://ui.adsabs.harvard.edu/abs/2015ApJ...801..113M}
}

@ARTICLE{Mowlavi2017,
       author = {{Mowlavi}, N. and {Lecoeur-Ta{\"\i}bi}, I. and {Holl}, B. and {Rimoldini}, L. and {Barblan}, F. and {Pr{\v{s}}a}, A. and {Kochoska}, A. and {S{\"u}veges}, M. and {Eyer}, L. and {Nienartowicz}, K. and {Jevardat}, G. and {Charnas}, J. and {Guy}, L. and {Audard}, M.},
        title = "{Gaia eclipsing binary and multiple systems. Two-Gaussian models applied to OGLE-III eclipsing binary light curves in the Large Magellanic Cloud}",
      journal = {\aap},
         year = 2017,
        month = oct,
       volume = {606},
          eid = {A92},
        pages = {A92},
          doi = {10.1051/0004-6361/201730613},
archivePrefix = {arXiv},
       eprint = {1703.10597},
 primaryClass = {astro-ph.IM},
       adsurl = {https://ui.adsabs.harvard.edu/abs/2017A&A...606A..92M}
}

@ARTICLE{Skowron2021,
       author = {{Skowron}, D.~M. and {Skowron}, J. and {Udalski}, A. and {Szyma{\'n}ski}, M.~K. and {Soszy{\'n}ski}, I. and {Wyrzykowski}, {\L}. and {Ulaczyk}, K. and {Poleski}, R. and {Koz{\l}owski}, S. and {Pietrukowicz}, P. and {Mr{\'o}z}, P. and {Rybicki}, K. and {Iwanek}, P. and {Wrona}, M. and {Gromadzki}, M.},
        title = "{OGLE-ing the Magellanic System: Optical Reddening Maps of the Large and Small Magellanic Clouds from Red Clump Stars}",
      journal = {\apjs},
         year = 2021,
        month = feb,
       volume = {252},
       number = {2},
          eid = {23},
        pages = {23},
          doi = {10.3847/1538-4365/abcb81},
archivePrefix = {arXiv},
       eprint = {2006.02448},
 primaryClass = {astro-ph.SR},
       adsurl = {https://ui.adsabs.harvard.edu/abs/2021ApJS..252...23S}
}

@ARTICLE{Pawlak2016,
       author = {{Pawlak}, M. and {Soszy{\'n}ski}, I. and {Udalski}, A. and {Szyma{\'n}ski}, M.~K. and {Wyrzykowski}, {\L}. and {Ulaczyk}, K. and {Poleski}, R. and {Pietrukowicz}, P. and {Koz{\l}owski}, S. and {Skowron}, D.~M. and {Skowron}, J. and {Mr{\'o}z}, P. and {Hamanowicz}, A.},
        title = "{The OGLE Collection of Variable Stars. Eclipsing Binaries in the Magellanic System}",
      journal = {\actaa},
         year = 2016,
        month = dec,
       volume = {66},
       number = {4},
        pages = {421-432},
          doi = {10.48550/arXiv.1612.06394},
archivePrefix = {arXiv},
       eprint = {1612.06394},
 primaryClass = {astro-ph.SR},
       adsurl = {https://ui.adsabs.harvard.edu/abs/2016AcA....66..421P}
}

@ARTICLE{VanHamme1993,
       author = {{Van Hamme}, W.},
        title = "{New Limb-Darkening Coefficients for Modeling Binary Star Light Curves}",
      journal = {\aj},
         year = 1993,
        month = nov,
       volume = {106},
        pages = {2096},
          doi = {10.1086/116788},
       adsurl = {https://ui.adsabs.harvard.edu/abs/1993AJ....106.2096V}
}

@ARTICLE{Sen2022,
       author = {{Sen}, K. and {Langer}, N. and {Marchant}, P. and {Menon}, A. and {de Mink}, S.~E. and {Schootemeijer}, A. and {Sch{\"u}rmann}, C. and {Mahy}, L. and {Hastings}, B. and {Nathaniel}, K. and {Sana}, H. and {Wang}, C. and {Xu}, X.~T.},
        title = "{Detailed models of interacting short-period massive binary stars}",
      journal = {\aap},
         year = 2022,
        month = mar,
       volume = {659},
          eid = {A98},
        pages = {A98},
          doi = {10.1051/0004-6361/202142574},
archivePrefix = {arXiv},
       eprint = {2111.03329},
 primaryClass = {astro-ph.SR},
       adsurl = {https://ui.adsabs.harvard.edu/abs/2022A&A...659A..98S}
}

@ARTICLE{Siess2018,
       author = {{Siess}, L. and {Lebreuilly}, U.},
        title = "{Case A and B evolution towards electron capture supernova}",
      journal = {\aap},
         year = 2018,
        month = jun,
       volume = {614},
          eid = {A99},
        pages = {A99},
          doi = {10.1051/0004-6361/201732502},
archivePrefix = {arXiv},
       eprint = {1807.04008},
 primaryClass = {astro-ph.SR},
       adsurl = {https://ui.adsabs.harvard.edu/abs/2018A&A...614A..99S}
}

@ARTICLE{Eggleton1983,
       author = {{Eggleton}, P.~P.},
        title = "{Aproximations to the radii of Roche lobes.}",
      journal = {\apj},
         year = 1983,
        month = may,
       volume = {268},
        pages = {368-369},
          doi = {10.1086/160960},
       adsurl = {https://ui.adsabs.harvard.edu/abs/1983ApJ...268..368E}
}

@ARTICLE{Pols1994,
       author = {{Pols}, O.~R.},
        title = "{Case A evolution of massive close binaries: formation of contact systems and possible reversal of the supernova order}",
      journal = {\aap},
         year = 1994,
        month = oct,
       volume = {290},
        pages = {119-128},
       adsurl = {https://ui.adsabs.harvard.edu/abs/1994A&A...290..119P}
}

@ARTICLE{Zapartas2017,
       author = {{Zapartas}, E. and {de Mink}, S.~E. and {Izzard}, R.~G. and {Yoon}, S.-C. and {Badenes}, C. and {G{\"o}tberg}, Y. and {de Koter}, A. and {Neijssel}, C.~J. and {Renzo}, M. and {Schootemeijer}, A. and {Shrotriya}, T.~S.},
        title = "{Delay-time distribution of core-collapse supernovae with late events resulting from binary interaction}",
      journal = {\aap},
         year = 2017,
        month = may,
       volume = {601},
          eid = {A29},
        pages = {A29},
          doi = {10.1051/0004-6361/201629685},
archivePrefix = {arXiv},
       eprint = {1701.07032},
 primaryClass = {astro-ph.HE},
       adsurl = {https://ui.adsabs.harvard.edu/abs/2017A&A...601A..29Z}
}

@ARTICLE{SciPy,
  author  = {Virtanen, Pauli and Gommers, Ralf and Oliphant, Travis E. and
            Haberland, Matt and Reddy, Tyler and Cournapeau, David and
            Burovski, Evgeni and Peterson, Pearu and Weckesser, Warren and
            Bright, Jonathan and {van der Walt}, St{\'e}fan J. and
            Brett, Matthew and Wilson, Joshua and Millman, K. Jarrod and
            Mayorov, Nikolay and Nelson, Andrew R. J. and Jones, Eric and
            Kern, Robert and Larson, Eric and Carey, C J and
            Polat, {\.I}lhan and Feng, Yu and Moore, Eric W. and
            {VanderPlas}, Jake and Laxalde, Denis and Perktold, Josef and
            Cimrman, Robert and Henriksen, Ian and Quintero, E. A. and
            Harris, Charles R. and Archibald, Anne M. and
            Ribeiro, Ant{\^o}nio H. and Pedregosa, Fabian and
            {van Mulbregt}, Paul and {SciPy 1.0 Contributors}},
  title   = {{{SciPy} 1.0: Fundamental Algorithms for Scientific
            Computing in Python}},
  journal = {Nature Methods},
  year    = {2020},
  volume  = {17},
  pages   = {261--272},
  adsurl  = {https://rdcu.be/b08Wh},
  doi     = {10.1038/s41592-019-0686-2},
}

@misc{UWARCC,
  doi       =   {10.15786/M2FY47},
  url       =   {https://arccwiki.atlassian.net/wiki/spaces/DOCUMENTAT/pages/1683587073/Beartooth},
  author    =   {{University of Wyoming Advanced Research Computing Center}},
  title     =   {{UW ARCC MedicineBow High Performance Compute Cluster}},
  publisher =   {University of Wyoming},
  year      =   {2018}
}

@ARTICLE{Yang2013,
       author = {{Yang}, Yuan-Gui and {Dai}, Hai-Feng and {He}, Jia-Jia and {Zhang}, Jia and {Ding}, Wei},
        title = "{New Photometric Models of Four Near-Contact Binaries: EP Cassiopeiae, AK Canis Minoris, FG Geminorum, and DF Puppis}",
      journal = {\pasj},
         year = 2013,
        month = apr,
       volume = {65},
          eid = {45},
        pages = {45},
          doi = {10.1093/pasj/65.2.45},
       adsurl = {https://ui.adsabs.harvard.edu/abs/2013PASJ...65...45Y}
}

@ARTICLE{Malkov2007,
       author = {{Malkov}, O. Yu.},
        title = "{Mass-luminosity relation of intermediate-mass stars}",
      journal = {\mnras},
         year = 2007,
        month = dec,
       volume = {382},
       number = {3},
        pages = {1073-1086},
          doi = {10.1111/j.1365-2966.2007.12086.x},
       adsurl = {https://ui.adsabs.harvard.edu/abs/2007MNRAS.382.1073M}
}

@ARTICLE{vanRensbergen2021,
       author = {{van Rensbergen}, Walter and {de Greve}, Jean-Pierre},
        title = "{On the Modeling of Algol-Type Binaries}",
      journal = {Galaxies},
         year = 2021,
        month = mar,
       volume = {9},
       number = {1},
          eid = {19},
        pages = {19},
          doi = {10.3390/galaxies9010019},
       adsurl = {https://ui.adsabs.harvard.edu/abs/2021Galax...9...19V}
}

@ARTICLE{Wang2022,
       author = {{Wang}, Z.~H. and {Zhu}, L.~Y. and {Yue}, Y.~F.},
        title = "{Evolutionary inference and statistical constraints on Algols including SD2-type near contact binaries}",
      journal = {\mnras},
         year = 2022,
        month = mar,
       volume = {511},
       number = {1},
        pages = {488-500},
          doi = {10.1093/mnras/stac037},
       adsurl = {https://ui.adsabs.harvard.edu/abs/2022MNRAS.511..488W}
}

@ARTICLE{Abt2002,
       author = {{Abt}, Helmut A. and {Levato}, Hugo and {Grosso}, Monica},
        title = "{Rotational Velocities of B Stars}",
      journal = {\apj},
         year = 2002,
        month = jul,
       volume = {573},
       number = {1},
        pages = {359-365},
          doi = {10.1086/340590},
       adsurl = {https://ui.adsabs.harvard.edu/abs/2002ApJ...573..359A}
}

@ARTICLE{Moe2013,
       author = {{Moe}, Maxwell and {Di~Stefano}, Rosanne},
        title = "{The Close Binary Properties of Massive Stars in the Milky Way and Low-metallicity Magellanic Clouds}",
      journal = {\apj},
         year = 2013,
        month = dec,
       volume = {778},
       number = {2},
          eid = {95},
        pages = {95},
          doi = {10.1088/0004-637X/778/2/95},
archivePrefix = {arXiv},
       eprint = {1309.3532},
 primaryClass = {astro-ph.SR},
       adsurl = {https://ui.adsabs.harvard.edu/abs/2013ApJ...778...95M}
}

@ARTICLE{Negu2018,
       author = {{Negu}, S.~H. and {Tessema}, S.~B.},
        title = "{Statistical analysis of Algol-type eclipsing binaries with stable mass transfer}",
      journal = {Astronomische Nachrichten},
         year = 2018,
        month = nov,
       volume = {339},
       number = {709},
        pages = {709-717},
          doi = {10.1002/asna.201813533},
       adsurl = {https://ui.adsabs.harvard.edu/abs/2018AN....339..709N}
}

@ARTICLE{Rucinski1969,
       author = {{Ruci{\'n}ski}, S.~M.},
        title = "{The Proximity Effects in Close Binary Systems. II. The Bolometric Reflection Effect for Stars with Deep Convective Envelopes}",
      journal = {\actaa},
         year = 1969,
        month = jan,
       volume = {19},
        pages = {245},
       adsurl = {https://ui.adsabs.harvard.edu/abs/1969AcA....19..245R}
}

@ARTICLE{Deschamps2013,
       author = {{Deschamps}, R. and {Siess}, L. and {Davis}, P.~J. and {Jorissen}, A.},
        title = "{Critically-rotating accretors and non-conservative evolution in Algols}",
      journal = {\aap},
         year = 2013,
        month = sep,
       volume = {557},
          eid = {A40},
        pages = {A40},
          doi = {10.1051/0004-6361/201321509},
archivePrefix = {arXiv},
       eprint = {1306.1348},
 primaryClass = {astro-ph.SR},
       adsurl = {https://ui.adsabs.harvard.edu/abs/2013A&A...557A..40D}
}

@ARTICLE{Naze2025,
       author = {{Naz{\'e}}, Ya{\"e}l and {Rauw}, Gregor and {Ko{\l}aczek-Szyma{\'n}ski}, Piotr A. and {Britavskiy}, Nikolay and {Labadie-Bartz}, Jonathan},
        title = "{A family of binaries with an extreme mass ratio}",
      journal = {\aap},
         year = 2025,
        month = nov,
       volume = {703},
          eid = {A239},
        pages = {A239},
          doi = {10.1051/0004-6361/202556441},
archivePrefix = {arXiv},
       eprint = {2510.15393},
 primaryClass = {astro-ph.SR},
       adsurl = {https://ui.adsabs.harvard.edu/abs/2025A&A...703A.239N}
}

\end{document}